\documentclass[preprint,showpacs,preprintnumbers,amsmath,amssymb]{revtex4}

\usepackage{graphicx}% Include figure files
\usepackage{dcolumn}% Align table columns on decimal point
\usepackage{bm}% bold math

\begin{document}

\preprint{APS/123-QED}

\title{A Statistical Theory of Homogeneous Isotropic Turbulence}% Force line breaks with \\

\author{Nicola de Divitiis}
 \altaffiliation[ ]{via Eudossiana, 18,  00184, Rome}%Lines break automatically or can be 
 \email{dedivitiis@dma.dma.uniroma1.it}
\affiliation{%
Department of Mechanics and Aeronautics\\
University "La Sapienza", Rome, Italy  %\textbackslash\textbackslash
}%

\date{\today}

\begin{abstract}
The present work proposes a theory of isotropic and homogeneous turbulence for incompressible fluids, which assumes that the turbulence is due to the bifurcations associated to the velocity field.
The theory is formulated using a representation of the fluid motion which is more general than the classical Navier-Stokes equations, where the fluid state variables are expressed in terms of the referential coordinates.

The theory is developed according to the following four items:
1) Study of the route toward the turbulence through the bifurcations analysis of the kinematic equations.
2) Referential description of the motion and calculation of  the velocity fluctuation
using the Lyapunov analysis of the local deformation.
3) Study of the mechanism of the energy cascade from large to small scales 
through the Lyapunov analysis of the relative kinematics equations of motion. 
4) Determination of the statistics of the velocity difference with the Fourier analysis.
Each item contributes to the formulation of the theory.

The theory gives the connection between number of bifurcations, scales and
Reynolds number at the onset of the turbulence and supplies an explanation for the mechanism of the energy cascade which leads to the closure of the von K\'arm\'an-Howarth equation.
The theory also gives the statistics of the velocity difference fluctuation and permits 
the calculation of its PDF.

The presented results show that the proposed theory describes quite well the properties of the isotropic turbulence.
\end{abstract}

\pacs{Valid PACS appear here}% PACS, the Physics and Astronomy
\maketitle

\newcommand{\no}{\noindent}
\newcommand{\be}{\begin{equation}}
\newcommand{\ee}{\end{equation}}
\newcommand{\bea}{\begin{eqnarray}}
\newcommand{\eea}{\end{eqnarray}}
\newcommand{\bc}{\begin{center}}
\newcommand{\ec}{\end{center}}

\newcommand{\calr}{{\cal R}}
\newcommand{\calv}{{\cal V}}

\newcommand{\bff}{\mbox{\boldmath $f$}}
\newcommand{\bfg}{\mbox{\boldmath $g$}}
\newcommand{\bfh}{\mbox{\boldmath $h$}}
\newcommand{\bfi}{\mbox{\boldmath $i$}}
\newcommand{\bfm}{\mbox{\boldmath $m$}}
\newcommand{\bfp}{\mbox{\boldmath $p$}}
\newcommand{\bfr}{\mbox{\boldmath $r$}}
\newcommand{\bfu}{\mbox{\boldmath $u$}}
\newcommand{\bfv}{\mbox{\boldmath $v$}}
\newcommand{\bfx}{\mbox{\boldmath $x$}}
\newcommand{\bfy}{\mbox{\boldmath $y$}}
\newcommand{\bfw}{\mbox{\boldmath $w$}}
\newcommand{\bfk}{\mbox{\boldmath $\kappa$}}

\newcommand{\bfA}{\mbox{\boldmath $A$}}
\newcommand{\bfD}{\mbox{\boldmath $D$}}
\newcommand{\bfI}{\mbox{\boldmath $I$}}
\newcommand{\bfL}{\mbox{\boldmath $L$}}
\newcommand{\bfM}{\mbox{\boldmath $M$}}
\newcommand{\bfS}{\mbox{\boldmath $S$}}
\newcommand{\bfT}{\mbox{\boldmath $T$}}
\newcommand{\bfU}{\mbox{\boldmath $U$}}
\newcommand{\bfX}{\mbox{\boldmath $X$}}
\newcommand{\bfY}{\mbox{\boldmath $Y$}}
\newcommand{\bfK}{\mbox{\boldmath $K$}}

\newcommand{\bfrho}{\mbox{\boldmath $\rho$}}
\newcommand{\bfchi}{\mbox{\boldmath $\chi$}}
\newcommand{\bfphi}{\mbox{\boldmath $\phi$}}
\newcommand{\bfPhi}{\mbox{\boldmath $\Phi$}}
\newcommand{\bflambda}{\mbox{\boldmath $\lambda$}}
\newcommand{\bfxi}{\mbox{\boldmath $\xi$}}
\newcommand{\bfLambda}{\mbox{\boldmath $\Lambda$}}
\newcommand{\bfPsi}{\mbox{\boldmath $\Psi$}}
\newcommand{\bfomega}{\mbox{\boldmath $\omega$}}
\newcommand{\bfeps}{\mbox{\boldmath $\varepsilon$}}
\newcommand{\bfkappa}{\mbox{\boldmath $\kappa$}}
\newcommand{\itPsi}{\mbox{\it $\Psi$}}
\newcommand{\itPhi}{\mbox{\it $\Phi$}}
\newcommand{\bint}{\mbox{ \int{a}{b}} }
\newcommand{\ds}{\displaystyle}
\newcommand{\Sum}{\Large \sum}

\section{\bf Introduction \label{s1} }

This work presents a theory of isotropic and homogeneous turbulence 
for an incompressible fluid formulated for an infinite fluid domain.
The theory is mainly  motivated by the fact that in turbulence the fluid kinematics 
is subjected to bifurcations \cite{Landau44} and exhibits a chaotic 
behavior and huge mixing \cite{Ottino90}, resulting to be much more rapid than the fluid 
state variables.
This characteristics implies that the accepted kinematical hypothesis for deriving the Navier-Stokes equations could require the consideration of very small length scales and times for describing the fluid motion \cite{Truesdell77} and therefore a very large 
number of degrees of freedom. 
To avoid the difficulties arising from the consideration of these small scales,
the referential description of motion is adopted, where the fluid state
variables are expressed in terms of the so called referential coordinates which coincide
with the material coordinates for a given fluid configuration \cite{Truesdell77}.
\\
The other very important subjects of the turbulence are the non-gaussian statistics of the velocity difference and the mechanism of the kinetic energy cascade.
This latter is directly related to the relative motion of a pair of fluid particles \cite{Richardson26, Kolmogorov41, Karman38, Batchelor53} and
is responsible for the shape of the developed energy spectrum.

For these reasons the present theory is based on: 

\begin{enumerate}
\item Landau hypothesis, following which the turbulence is caused by the  bifurcations of the velocity field \cite{Landau44}.

\item Referential description of motion, where velocity field and stress tensor are mapped with respect to the referential coordinates \cite{Truesdell77}.

\item Study of the energy cascade through Lyapunov analysis of the relative kinematics.

\item Statistical analysis of the velocity difference fluctuations.
\end{enumerate}

In the first part of the work, the road toward the turbulence is studied 
through the bifurcations analysis of the kinematic equations. These bifurcations arise from the mathematical structure of the
velocity field, where the Reynolds number plays the role of the "control parameter".
This analysis supplies the connection between number of bifurcations and the critical  Reynolds number for isotropic turbulence, showing that the length scales are continuously distributed and that each of them is important for the description of the motion.

In the second part, the momentum equations are formulated according to the referential representation of motion,  whereas the kinematics of the local deformation is studied with the Lyapunov theory.
The fluid motion is described adopting the referential configuration which corresponds to the fluid placement at the onset of  this fluctuation.
This  choice allows the velocity fluctuations to be analytically expressed through the
Lyapunov analysis of the kinematics of the fluid deformation.

The third part deals with the relative kinematic between two trajectories, which is also analyzed with the Lyapunov theory. This analysis gives an explanation of the mechanism of kinetic energy transfer between length scales and leads to the closure of the von K\'arm\'an-Howarth equation \cite{Karman38} (see Appendix), where the unknown function $K(r)$, which represents the inertia forces, is here expressed in terms of the longitudinal correlation function. The obtained expression of $K(r)$ satisfies the conservation law which states that the inertia forces only transfer the kinetic energy 
\cite{Karman38, Batchelor53}.

To complete the theory, the statistics of velocity difference is studied through
the Fourier analysis of the velocity fluctuations. 
An analytical expression for the velocity difference and for its PDF
is obtained in case of isotropic turbulence.
This expression incorporates an unknown function, related to the skewness, 
which is immediately identified through the obtained expression of $K(r)$.

Finally, the several results obtained with this theory are compared with the data
existing in the literature, indicating that the proposed theory adequately describes the
various properties of the turbulence.
\begin{figure}[b]
\centering
\vspace{-0.mm}
\hspace{-0.mm}
\includegraphics[width=0.450\textwidth]{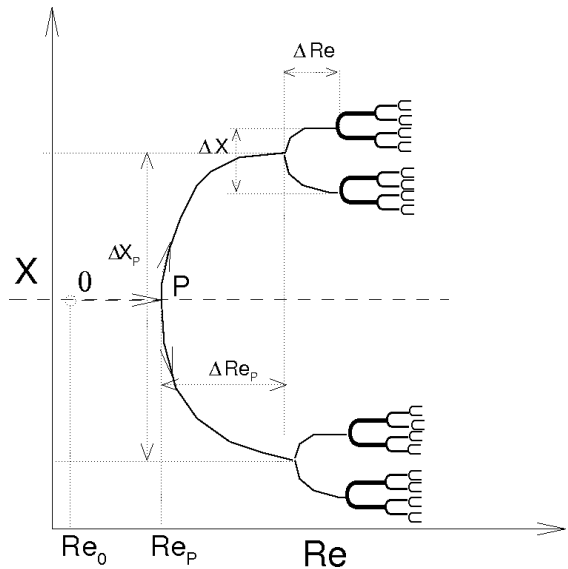}
\caption{Map of the bifurcations.}
\label{figura_1}
\end{figure}

\section{\bf Bifurcation Analysis of the Kinematic Equations  \label{s2}}

In this session, the route toward the turbulence is studied through
the analysis of the bifurcations of the kinematic equations.
%This analysis provides the connection between bifurcations,
%length scales and critical Reynolds number.
To analyze this question, a viscous and incompressible fluid in the infinite domain is considered, whose kinematic equations are
\bea
\ds \frac{d{\bf x}}{dt} = {\bf u}({\bf x}, t  ; Re)
\label{0_00}
\label{0_0}
\eea
where $\bf x$ and $Re$ are the position and Reynolds number, 
whereas ${\bf u}({\bf x}, t  ; Re)$ is a single realization of the ensemble 
of the velocity fields, written in the reference frame $\Re$, 
which satisfies the Navier-Stokes equations
\bea
\begin{array}{l@{\hspace{-1.cm}}l}
\nabla \cdot {\bf u} = 0 \\\\
\ds \frac{\partial {\bf u}}{\partial t} +{\bf u} \nabla {\bf u} 
\ds +\frac{\nabla p}{\rho} - \nu \nabla^2 {\bf u} = 0
\end{array}
\label{N-S}
\eea
$\rho$ and $\nu$ are, respectively, density and kinematic viscosity
whereas $p$ is the fluid pressure which can be eliminated by taking the divergence of the momentum equation \cite{Batchelor53}
\bea
\ds \frac{\nabla^2 p}{\rho} + \nabla {\bf u} : \nabla {\bf u} = 0
\label{inc}
\eea

Now, let consider an assigned velocity field at a given time, and the fixed points ${\bf X}$ of 
Eq. (\ref{0_0}) which satisfy to $\ds {d{\bf X}}/{dt}$ = 0.
Increasing the Reynolds number, ${\bf X}$ will vary according to Eq. (\ref{0_0}),
which can be solved by the continuation method \cite{Guckenheimer90, Kuznetsov04}
\bea
{\bf X}  = {\bf X}_0 - \int_{Re_0}^{Re} 
%\left( \frac{\partial{\bf u}  }{\partial {\bf x}
% }\right)^{-1}
\nabla {\bf u}^{-1}
\frac{\partial{\bf u}  }{\partial Re} \ dRe
\label{bif_map}
\eea
where ${\bf X}_0$ is the fixed point calculated at $Re=Re_0$.
The Reynolds number influences the mathematical structure of Eq. (\ref{0_0})
through the Navier-Stokes equations in such a way that, for small $Re$,
the viscosity forces which are stronger than the inertia ones,
make $\bf u$ an almost smooth function of $\bf X$. 
When the Reynolds number increases, as long as the Jacobian 
$\nabla {\bf u}$ is nonsingular, $\bf X$ exhibits smooth
variations with $Re$, whereas at a certain $Re$, this Jacobian becomes singular
due to the higher inertia-viscous forces ratio, resulting
$\det \left( \nabla {\bf u}\right) $ = 0. 
This can correspond to the first bifurcation, where at least one of the eigenvalues of 
$\nabla {\bf u}$ crosses the imaginary axis and $\bf X$ 
appears  to be discontinuous with respect to $Re$ \cite{Guckenheimer90, Kuznetsov04}.
Increasing again the Reynolds number, $\bf X$ will show smooth variations until to the next bifurcation.

Figure 1 shows a scheme of bifurcations, where the component $X$ of $\bf X$
is reported in terms of Reynolds number. Starting from $Re_0$, the diagram is regular, 
until to $Re_P$, where the first bifurcation determines two branches, whose distance  $\Delta X_P$ is measured at the next bifurcation.
For each bifurcation, $\Delta X$ gives a length scale of the velocity field at the current Reynolds number, whereas $\Delta Re$ represents the distance between two successive bifurcations.
After $P,$ Eq. (\ref{bif_map}) does not indicate which of the two possible branches
the system will choose, thus a bifurcation causes a lost of informations with respect to the initial data \cite{Prigogine94}. Therefore, the fluctuations  are important
for the choice of the branch that the system will follow \cite{Prigogine94}. 

Further increments of $Re$ cause an increment of the number of bifurcations whose scaling laws are described by the two successions \cite{Feigenbaum78, Kuznetsov04}
\bea
\alpha_n = \frac{\Delta X_n}{\Delta X_{n+1}},  \ \
\delta_n = \frac{\Delta Re_n}{\Delta Re_{n+1}}
\eea
For $Re\rightarrow\infty$, the convergence of $\alpha_n$ and $\delta_n$ is not granted in general, whereas for period-doubling bifurcations, these admit the following  limits \cite{Feigenbaum78}
\bea
\alpha = \lim_{Re \rightarrow \infty} \vert \alpha_n \vert  = 2.502...\ \
\delta = \lim_{Re \rightarrow \infty} \vert \delta_n \vert  = 4.669...
\label{Feigen_c}
\eea
These are the famous Feigenbaum numbers, which are two universal constants, independent on the mathematical details of the period-doubling bifurcations.
For bifurcations of other kind, $\alpha_n$ and $\delta_n$ can converge to different values
or can oscillate around to average values.

In the present analysis, the length scales $l_n \equiv \Delta X_n$ are assumed to
be expressed by the asymptotic approximation 
\bea
l_n = \frac{l_{1}} {\alpha^{n-1}}
\label{scales0}
\eea
Equation (\ref{scales0}) supplies the length scales in terms
of the numbers of the bifurcations encountered along a given
path of fixed points, where $\alpha$ is the Feigenbaum constant given by Eq. (\ref{Feigen_c}) and $l_1$ represents the maximum length scale.
According to \cite{Ruelle71, Feigenbaum78, Pomeau80, Eckmann81}, 
the bifurcations generate a route toward the chaos which depends on $n$.
As long as $n \le$ 2, each bifurcation adds a new frequency into the power spectrum of 
$\bf u$ and this corresponds to limit cycles or quasi periodic motions, whereas
for $n \ge$ 3, the situation drastically changes, since $\bf u$ 
exhibits more numerous frequencies and this generates chaotic motion 
\cite{Ruelle71, Feigenbaum78}.
This occurs for a single realization of the ensemble of the velocity field.
The fluctuations of ${\bf u} ({\bf x}, t)$ will cause further variations of the several scales 
$l_n$ in Eq. (\ref{scales0}), thus the bifurcations maps will be more complicated than Fig. \ref{figura_1}, and the recognizing the diverse scales and bifurcations could not be  possible.
This is a scenario with continuously distributed length scales, where all of them are important for describing the fluid motion.

\subsection{\bf  Critical Reynolds number \label{s3} }

Equation (\ref{scales0}) describes the route toward the chaos and is assumed to be valid until the onset of the turbulence. 
In this situation the minimum for $l_n$ can not be less than the dissipation length or Kolmogorov scale $\ds \ell = (\nu^3/ \varepsilon)^{1/4}$ \cite{Landau44}, where
 $\varepsilon$ is the energy dissipation rate (see Appendix), whereas $l_1$ gives a good
 estimation of the correlation length of the phenomenon \cite{Guckenheimer90, Prigogine94}
 which, in this case is the Taylor scale $\lambda_T$.
Thus, $\ell < l_n < \lambda_T$, and
\bea
\ds \ell = \frac{\lambda_T}  {\alpha^{N-1}}
\label{scales01}
\eea
where $N$ is the number of bifurcations at the beginning of the turbulence.

Equation (\ref{scales01}) gives the connection between 
the critical Reynolds number and number of bifurcations.
In fact, the characteristic Reynolds numbers associated to the scales 
$\ell$ and $\lambda_T$ are $R_K = \ell u_K/\nu \equiv$ 1 and
 $R_{\lambda} = \lambda_T u/\nu$, respectively, where 
$\ds u_K = (\nu \varepsilon )^{1/4}$ is characteristic velocity at the Kolmogorov scale,
 and $u =\sqrt{\left\langle u_i u_i \right\rangle/3}$ is the velocity standard
deviation \cite{Batchelor53}.
For isotropic turbulence, these scales are linked each other by \cite{Batchelor53}
\bea
\ds {\lambda_T}/{\ell} = 15^{1/4} \sqrt{R_\lambda}
\label{scales}
\eea
In view of Eq.(\ref{scales01}), this ratio can be also expressed through $N$.
\bea
\alpha^{N-1} = 15^{1/4} \sqrt{R_\lambda}
\label{scales1}
\eea
The value $R_\lambda \simeq$ 1.613 obtained for $N=$ 2 is not compatible with $\lambda_T$
which is the correlation scale, while the result $R_\lambda \simeq$ 10.12, calculated for
$N=$ 3, is an acceptable minimum value for $R_\lambda$. This result agrees with the various scenarios describing the roads to the turbulence  \cite{Ruelle71, Feigenbaum78, Pomeau80, Eckmann81}, and with the diverse 
experiments \cite{Gollub75, Giglio81, Maurer79} which state that the turbulence begins 
for $N \ge 3$.
Of course, this minimum value for $R_\lambda$ is the result of the assumptions $\alpha \simeq$ 2.502,
$l_1 \simeq \lambda_T$, $l_N \simeq \ell$ and of the asymptotic approximation (\ref{scales0}).

\begin{figure}[b]
\vspace{-0. mm}
\hspace{-0. mm}
	\centering
         \includegraphics[width=0.45\textwidth]{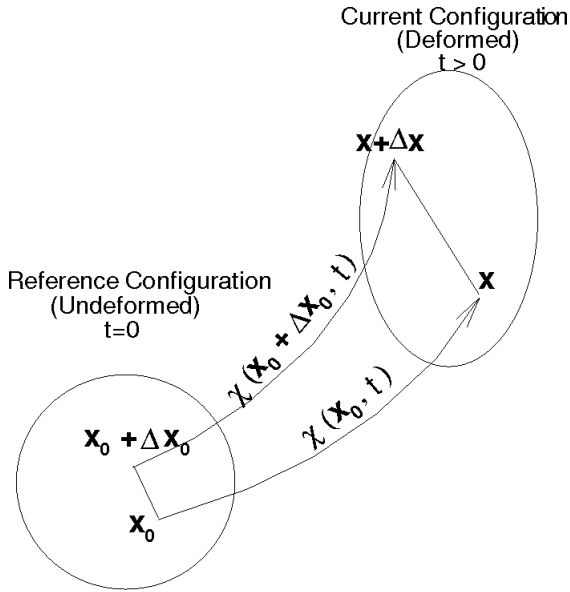}
% trasp.jpg: 100dpi, width=7.62cm, height=5.38cm, bb=0 0 300 212
\caption{Referential description of the kinematics of deformation.}
\label{figura_1ab}
\end{figure}

\section{\bf referential description of motion. Velocity fluctuation  \label{s3a} }

Now, we present a formulation of the fluid equations of motion which is based on the
referential description of the motion. This formulation is more general than the classical  Navier-Stokes equations and is capable to take into account the effects of the fluid kinematics which can be much faster than the fluid state variables. 
This description of motion allows to calculate the velocity fluctuation through the Lyapunov analysis of the local deformation.

This representation of motion is based on the fact that a given fluid property
$\Omega$ is an explicit function of the referential displacement ${\bf x}_0$
and of the time \cite{Truesdell77}, i.e.
\bea
\Omega = \Omega ({\bf x}_0, t ) 
\eea
The referential displacement coincides with the material position
for a given fluid configuration, thus ${\bf x}_0$ plays the role of the label which
identifies the specific fluid particle  \cite{Truesdell77}.
Since any fluid motion has infinitely many different referential descriptions
which are equally valid \cite{Truesdell77}, it is convenient to choose 
the referential configuration corresponding to the fluid placement at the onset of the deformation (see Fig. \ref{figura_1ab}).
According to Truesdell \cite{Truesdell77}, $\Omega ({\bf x}_0, t )$ and its 
derivatives with respect to ${\bf x}_0$ are supposed to be smooth functions 
of $t$ and ${\bf x}_0$.
Hence, if ${\bf x} = \bfchi({\bf x}_0, t)$ represents the fluid motion,
$\Omega$ is expressed in terms of the geometrical position ${\bf x}$,
through the inverse of $\bfchi$, ${\bf x}_0$ = $\bfchi^{-1} ({\bf x}, t)$ 
\bea
\Omega ({\bf x}, t ) = \Omega (\bfchi^{-1}({\bf x}, t), t )
\label{Ox}
\eea
and its derivative with respect to $\bf x$ is 
\bea
\frac{\partial \Omega}{\partial {\bf x}} = 
\frac{\partial \Omega}{\partial {\bf x}_0}    \frac{\partial {\bf x}_0} {\partial {\bf x}}
\label{Oxx}
\eea
The bifurcations of Eq. (\ref{0_00}) make $\bfchi$ a singular transformation,
thus, in proximity of a bifurcation, $\ds {\partial \Omega}/{\partial {\bf x}}$  varies much more quickly than $\ds {\partial \Omega}/{\partial {\bf x}_0}$ because of the local stretching $\ds  {\partial {\bf x}} /{\partial {\bf x}_0}$, which is here calculated with the Lyapunov theory, as
\bea
\frac{\partial {\bf x}}{\partial {\bf x}_0} \approx {\mbox e}^{\Lambda t} 
\label{stretch}
\eea
where $\Lambda = \max (\Lambda_1, \Lambda_2, \Lambda_3 )$ is the maximal Lyapunov exponent and $\Lambda_i$, $(i = 1, 2, 3)$ are the Lyapunov exponents.
Due to the incompressibility, $\Lambda_1 + \Lambda_2 + \Lambda_3 =$ 0,
thus,  $\Lambda >0$.

The velocity fluctuation of the particle ${\bf x}_0$ -or Lagrangian fluctuation- is calculated using the momentum equations, where stress tensor and velocity field are mapped with respect to the referential coordinates at the beginning of the deformation 
\bea
\ds \left( \frac{ \partial {u}_k}{\partial t} \right)_{{\bf x}_0} 
%({\bf x}_0, t) 
 =
\frac{1}{\rho}
 \frac{\partial T_{k h}}  {\partial x_{j 0}} ({\bf x}_0, t)
  \frac{\partial x_{j 0}}  {\partial x_h} ({\bf x}_0, t)
\label{q_dot}
\eea
where $\left( \partial {u}_k / \partial t \right)_{{\bf x}_0}$ is the acceleration 
of the particle ${\bf x}_0$, whereas $T_{k h}$ represents the stress
tensor 
\bea
T_{h k} = 
\ds - {p} \delta_{h k} + 
\nu \rho \left(  \frac{\partial u_h}{\partial x_k} + \frac{\partial u_k}{\partial x_h} \right) 
\eea
Note that, Eq. (\ref{q_dot}) is more general than the classical Navier-Stokes
equations, since it can be applied to fluid particles which exhibit non-smooth
displacements and irregular boundaries \cite{Truesdell77}, as in the present case.
Since ${\partial {\bf x}}/{\partial {\bf x}_0}$ is much more rapid than 
$\ds {\partial T_{k h}}/{\partial x_{j 0}}$, this fluctuation is calculated integrating Eq. (\ref{q_dot}) from $t=0$ to $\infty$,
considering ${\partial T_{k h}}/{\partial x_{j 0}}$ constant with respect to 
${\partial {\bf x}}/{\partial {\bf x}_0}$, i.e.
\bea
\begin{array}{l@{\hspace{0cm}}l}
\ds u_k ({\bf x}_0)\approx 
\frac{1}{\Lambda}  
\left( 
\ds - \frac{1}{\rho} \frac{\partial p}{\partial x_k} + \nu \nabla^2 u_k 
\right)
\ds = \frac{1}{\Lambda}  \left( \frac{\partial u_k} {\partial t} \right) _{{\bf x}_0}
\end{array}
\label{fluc_v2_0}
\eea
The velocity fluctuation in a fixed point of space $\bf x$
-or Eulerian fluctuation- is calculated  taking into account the expression of the Eulerian time derivative of $u_k$, which is \cite{Truesdell77}
\bea
\ds \left( \frac{\partial {u}_k}{\partial t} \right)_{{\bf x}}  =
\ds \left( \frac{\partial {u}_k}{\partial t} \right)_{{\bf x}_0} -
\ds \frac{\partial {u}_k}{\partial x_{h 0}}  ({\bf x}_0, t) 
\frac{\partial {x}_{h 0}}{\partial x_{j}}   u_j
\eea
Therefore, this velocity fluctuation is  
\bea
\begin{array}{l@{\hspace{0cm}}l}
\ds u_k ({\bf x})\approx 
%\frac{1}{\Lambda}  
%\left( -\frac{\partial u_k}{\partial x_j} u_j 
%\ds - \frac{1}{\rho} \frac{\partial p}{\partial x_k} + 
%\nu \nabla^2 u_k 
%\right) \\\\
\ds  \frac{1}{\Lambda} \left( \frac{\partial u_k} {\partial t} \right)_{{\bf x}}
\end{array}
\label{fluc_v2}
\eea

These velocity fluctuations, which stem from the bifurcations of the velocity
field, do not modify the average values of the momentum and of the kinetic energy of
fluid.

\section{\bf Lyapunov analysis of the relative kinematics \label{s5}}

\begin{figure}
\vspace{-0.mm}
\centering
 \includegraphics[width=0.4\textwidth]{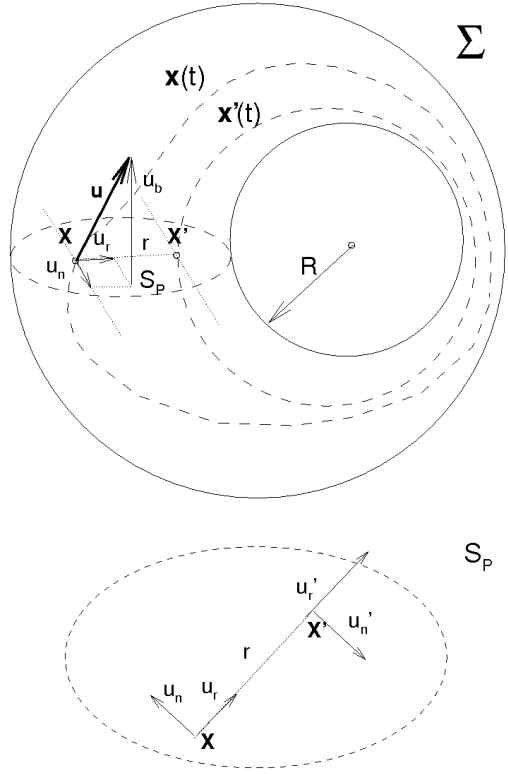}
\vspace{-0.mm}
\caption{Scheme of the relative kinematics of two fluid particles}
\vspace{-0.mm}
\label{figura_1a}
\end{figure}

In order to investigate the mechanism of the energy cascade, the properties of the relative kinematic equations are here studied with the Lyapunov analysis. 
These equations are 
\bea
\begin{array}{l@{\hspace{+0.2cm}}l}
\ds \frac{d {\bf x}}{d t} = {\bf u} ({\bf x}, {t}),  \ \  \frac{d {\bf x}'}{d t}
 = {\bf u} ({\bf x}', {t})
\end{array}
\label{k_1}
\label{k_0}
\eea 
where ${\bf u}({\bf x}, t)$ = $(u_1, u_2, u_3)$, 
${\bf u} ({\bf x}', t)$ $\equiv$ ${\bf u}'= (u_1', u_2', u_3')$, whereas 
$u_i$ and $u_i'$ 
are the  velocity components expressed in the reference frame $\Re$.
Since the bifurcations do not modify the total momentum and kinetic energy, the
solutions of Eq. (\ref{k_1}) preserve these quantities.
With reference to Fig. \ref{figura_1a}, these solutions correspond to the paths,
${\bf x}(t)$ and ${\bf x}'(t)$, located into a material volume $\Sigma (t)$ 
which changes its geometry according to the fluid motion \cite{Lamb45},
whereas its volume remains unaltered.
This is a toroidal volume, where $S_p$ and $R$ are, respectively, the poloidal surface and the toroidal dimension of $\Sigma$, whereas ${\bf X}$ and ${\bf X}'$ are the intersections of ${\bf x}(t)$ and ${\bf x}'(t)$ with $S_p$, where 
$r = \vert {\bf X}'-{\bf X} \vert$ is the poloidal dimension, thus $S_p \approx r^2$. 
The velocity difference components $\Delta u_n \equiv u_n'-u_n$ and 
$\Delta u_r \equiv u_r'-u_r$ lay on $S_p$ and are normal and parallel to $r$, respectively, whereas $u_b$ is the average velocity component along the direction normal to $S_p$.
The equations describing the evolution of these quantities preserve the
volume and the momentum of $\Sigma$. These can be written as 
\bea
\begin{array}{l@{\hspace{+0.0cm}}l}
\ds \frac{d}{dt} \left( S_p R  \right) = 0 
\end{array}
\label{laws0}
\eea
\bea
\begin{array}{l@{\hspace{+0.0cm}}l}
\ds \frac{d}{dt} \left( \Delta u_n^2 S_p \right) = 0 \\\\
\ds \frac{d}{dt} \left( u_b R  \right) = 0
\end{array}
\label{laws1}
\eea
Equations  (\ref{laws0}) and  (\ref{laws1}) represent, respectively, the continuity equation, and the momentum equations according to the third Helmholtz theorem on the vorticity 
\cite{Truesdell77, Lamb45}.

The Lyapunov analysis, applied to Eqs. (\ref{laws0}) and (\ref{laws1}), states that  
$R \approx  R_0 \ \mbox{e}^{\lambda t}$, 
hence, Eqs. (\ref{laws0}) and (\ref{laws1}) become
\bea
\begin{array}{l@{\hspace{+0.0cm}}l}
\ds \frac{d \Delta u_r^2}{d t} = - \lambda \Delta u_r^2 \\\\
\ds \frac{d \Delta u_n^2}{d t} =  \lambda \Delta u_n^2 \\\\
\ds \frac{d u_b}{d t} =  - \lambda  u_b
\end{array}
\label{laws-lyap}
\eea
where $\lambda(r) >$0  is the maximal finite scale Lyapunov exponents 
associated to Eqs. (\ref{k_0}), with $\lambda(0) = \Lambda$. 
As the result,  $u_b \rightarrow 0$ and 
$\Delta {\bf u} \approx \Delta u_n \propto \mbox{e}^{\lambda/2 t}$.

Now, it is worth to remark that the following quantity 
\bea
%\ds \Upsilon \equiv \frac{d }{dt}\sum_{i= r, n, b}  u_i u_i'
\ds \Upsilon \equiv \frac{d }{dt} \left( {\bf u}\cdot {\bf u}' \right) 
\label{Upsilon}
\eea
expresses the transfer of the kinetic energy between the points ${\bf x}$ and ${\bf x}'$.
Its average $\langle \Upsilon \rangle$ is calculated on the ensemble of the diverse pairs of trajectories which pass through ${\bf X}$ and ${\bf X}'$ and which are contained into the various toroidal volumes.
This average is obtained from Eqs. (\ref{laws-lyap}), taking into account the
 homogeneity, the isotropy and the time independence upon the time of the average kinetic 
energy ( $\ds d  \langle {\bf u}\cdot {\bf u}  \rangle /dt$ = 0).
\bea
\langle \Upsilon \rangle =
\ds \langle \frac{d} {dt} \sum_{i= r, n, b}  u_i u_i' \rangle  =
 \lambda \ u^2 \ ( g - f ) 
\label{vv'}
\eea
$f$ and $g$ are longitudinal and lateral velocity correlation functions, that, because of the incompressibility, are related each other through Eq. (\ref{g}) 
(see Appendix). Thus, $\langle \Upsilon \rangle$ is 
\bea
\langle \Upsilon \rangle =
\frac{1}{2} u^2    \frac{\partial f}{\partial r} \ \lambda(r) r
\label{vv'1}
\eea
If $\Upsilon$ were an ergodic function, its average on the statistical 
ensemble should coincide with the average over time which in turn is 
equal to zero since $\Upsilon$ is the time derivative
of ${\bf u}\cdot {\bf u}'$. 
As the consequence, there would not be any transfer of energy between the 
parts of fluid.
\\
Therefore, the fluid incompressibility is a sufficient condition to state that $\Upsilon$ is
a non ergodic function, whose statistical average  is determined 
as soon as $\lambda$ is known. 
To calculate $\lambda$, it is convenient to express the velocity difference
$\Delta {\bf u} = {\bf u}({\bf x}', t) - {\bf u}({\bf x}, t)$ in the Lyapunov basis $E$
associated to Eqs. (\ref{k_1}), which is made by orthonormal vectors arising from Eqs. (\ref{k_1}) \cite{Christiansen97, Ershov98}.
The velocity difference expressed in $E$, 
$\Delta {\bf v} \equiv (v_1' -v_1, \ v_2' -v_2, \ v_3' -v_3 )$,  
satisfies the following equations, which hold for ${t \rightarrow \infty}$
\bea 
\begin{array}{l@{\hspace{+0.0cm}}l}
\ds v_i' -v_i   =  {\lambda}_ i \ \hat{r}_i, \ \ \ i = 1, 2, 3 
\end{array}
\label{Lyap}
\eea
where $\hat{r}_i$,  $v_i$ and  $v_i'$ 
are, respectively, the components of $\hat{\bf r} \equiv {\bf x}'-{\bf x}$, 
${\bf u} ({\bf x}, t) $ and ${\bf u} ({\bf x}', t)$ written in $E$.
Then, $\Delta u_r$ and $r$ can be expressed in terms of 
$\Delta {\bf v}$ and $\hat{\bf r}$ as
\bea
\begin{array}{l@{\hspace{+0.0cm}}l}
r \approx  {\bfxi} \cdot {\bf Q} \hat{\bf r},  \ \ \
\Delta u_r \approx  {\bfxi} \cdot   {\bf Q} \Delta {\bf v}
\end{array}
\label{rot_E_R}
\eea
Into Eqs. (\ref{rot_E_R}), ${\bf Q}$ is the fluctuating rotation matrix transformation from $E$ to $\Re$, and ${\bfxi} = ({\bf X}'-{\bf X})/\vert {\bf X}'-{\bf X} \vert $.
The standard deviation of $\Delta u_r$ is calculated from Eqs. (\ref{rot_E_R}), taking into 
account the isotropy and that $\Delta {\bf v} \approx \lambda  \hat{\bf r}$
\bea
\left\langle \Delta u_r^2 (r) \right\rangle = {\lambda}^2 r^2
\label{1A}
\eea 
This standard deviation can be also expressed through the longitudinal correlation function $f$
\bea
\langle \Delta u_r^2(r) \rangle = 2 u^2 (1-f(r))
\label{1B}
\eea
being $u$ the standard deviation of the longitudinal velocity.
The maximal Lyapunov exponent is calculated in function of $f$,
from Eqs. (\ref{1A}) and (\ref{1B})
\bea
\ds {\lambda} (r) = \frac{u}{r} \sqrt{2 \left( 1-f(r) \right) }
\label{lC}
\eea
Hence, substituting Eq. (\ref{lC}) into Eq. (\ref{vv'1}), one obtains
the expression of $\langle \Upsilon \rangle$ in terms of the longitudinal
correlation function
\bea
\langle \Upsilon \rangle =
u^3 \sqrt{\frac{1-f}{2}} \ \frac{\partial f}{\partial r} 
\label{vv'2}
\eea
where, thanks to the isotropy, $\langle \Upsilon \rangle$ is a function
of $r$ alone.

Equation (\ref{vv'2}) reflects the well known property 
of the inertia forces of transferring the kinetic energy \cite{Batchelor53}
between the several regions of the fluid domain.

\section{\bf Closure of the von K\'arm\'an-Howarth equation \label{s6} }

The closure of the von K\'arm\'an-Howarth equation is now
carried out using the previous Lyapunov analysis.
\\
The function $K(r)$ is defined through the following relation 
(see also Eq. (\ref{vk1}) in the Appendix)
\bea
\ds  \frac{\partial }{\partial {r}_k} 
\ds  \left\langle u_i  u_i' (u_k - u_k')  \right\rangle =
\frac{\partial K(r)}{\partial r } r + 3 K(r)
\label{vk1_0}  
\eea
The repeated indexes into Eq. (\ref{vk1_0}), $i$ and $k$, indicate the summations with respect to the same indexes.
In order to obtain the expression of $K(r)$, it is worth to remark the following identity
\bea
u_i  u_i' (u_k - u_k') = \Upsilon \ r_k - \frac{d}{dt} ( u_i  u_i' r_k ) 
\label{id10}
\eea
The average of Eq. (\ref{id10}) is calculated on the ensemble of the trajectories
passing through $\bf X$ and ${\bf X}'$.
It is supposed that the ergodic hypothesis holds for the last term at the 
right hand-side of  Eq. (\ref{id10}), thus this latter can be calculated through the 
average over time.
Since this term is the time derivative of $ u_i  u_i' r_k$, this gives null contribution.
Hence, accounting for the isotropy, one obtains 
\bea
\hspace{-2. mm}
\frac{\partial}{\partial r_k} \langle u_i  u_i' (u_k - u_k') \rangle=
\frac{\partial \langle \Upsilon \rangle}{\partial r } r + 3 \langle \Upsilon \rangle
\label{id1}
\eea
Comparing Eqs. (\ref{vk1_0}) and (\ref{id1}), and taking into account that
$K(0) = 0$ \cite{Batchelor53}, $K(r) \equiv \langle \Upsilon \rangle$, i.e.
\bea
\ds K(r) =  u^3 \sqrt{\frac{1-f}{2}} \ \frac{\partial f}{\partial r} 
\label{vk6}
\eea
Equation (\ref{vk6}) represents the proposed closure of the von K\'arm\'an-Howarth equation, and expresses the transfer of kinetic energy between the diverse fluid regions.
This is a kinematic mechanism, caused by the bifurcations cascades of Eq. (\ref{k_1}), 
which preserves total momentum and kinetic energy.
The analytical structure of Eq.(\ref{vk6}) states that this mechanism consists of a flow of the kinetic energy from large to small scales which only redistributes the kinetic energy between wavelengths. 

\bigskip

The skewness of $\Delta u_r$ is determined once $K(r)$ is known \cite{Batchelor53}.
This is
\bea
\ds H_3(r) = \frac{\left\langle \Delta u_r^3 \right\rangle} 
{\left\langle \Delta u_r^2\right\rangle^{3/2}} =
  \frac{6 k(r)}{\left( 2 (1 -f(r)  )   \right)^{3/2} }
\label{H_3_01}
\eea
The longitudinal triple correlation $k(r)$ is calculated by Eq. (\ref{kk}) (see Appendix).
Since $f$ and $k$ are, respectively, even and odd functions of $r$ with
 $f(0)$ = 1, $k(0) = k'(0)=k''(0)$ =0,   $ H_3(0)$ is given by 
 \bea
\ds H_3(0) = \lim_{r\rightarrow0} H_3(r) = \frac{k'''(0)}{(-f''(0))^{3/2}} 
\label{H_3_0}
\eea
where the apex denote the derivative with respect to $r$.
To obtain $H_3(0)$, observe that, near the origin, $K$ behaves as
\bea
\begin{array}{l@{\hspace{+0.cm}}l}
 \ds K = u^3 \sqrt{-f''(0)} f''(0) \frac{r^2}{2} + O(r^4)
\end{array}
\label{K0}
\eea
then, substituting Eq. (\ref{K0}) into Eq. (\ref{kk}) (see Appendix) and accounting for Eq. (\ref{H_3_0}), one obtains
\bea
\ds H_3(0)  = -\frac{3}{7} = -0.42857...
\label{sk0}
\eea
This value of  $H_3(0)$ is a constant of the present theory, 
which does not depend on the Reynolds number. 
This is in agreement with the several sources of data existing in the literature
such as \cite{Batchelor53, Tabeling96, Tabeling97, Antonia97} (and Refs. therein) and the knowledge of it gives the entity of the mechanism of energy cascade.

\section{\bf Statistical analysis of velocity difference \label{s4}}

Although the previous analysis 
leads to the closure of the von K\'arm\'an-Howarth equation, it does not give any information about the statistics of velocity difference
$\Delta {\bf u} ({\bf r}) \equiv {\bf u} ({\bf x}+{\bf r})-{\bf u} ({\bf x})$.

In this section, the statistical properties of $\Delta {\bf u} ({\bf r})$, are  investigated through the Fourier analysis of the velocity fluctuation given by Eq. (\ref{fluc_v2}).
This fluctuation is 
\bea
\ds {\bf u} =  
\sum_{\bfkappa} {\bf U} ({\bfkappa}) {\mbox e}^{i {\bfkappa}\cdot {\bf x}}  
\approx
 \frac{1}{\Lambda} \sum_{\bfkappa} \frac{{\bf \partial U}}{\partial t} ({\bfkappa}) 
 {\mbox e}^{i {\bfkappa}\cdot {\bf x}}  
\label{f1}
\eea
where ${\bf U}({\bfkappa})$ $\equiv$ $(U_1({\bfkappa}), U_2({\bfkappa}), U_3({\bfkappa}))$ are the components of velocity spectrum, which satisfy \cite{Batchelor53}
\bea
\begin{array}{l@{\hspace{0.cm}}l}
\ds  \frac{\partial  U_p ({\bfkappa})}{\partial t}  = - \nu k^2 U_p ({\bfkappa}) + \\\\
\ds i \sum_{\bf j} ( \frac{\kappa_p \kappa_q \kappa_r}{\kappa^2}   U_q({\bf j}) 
U_r({\bfkappa} -{\bf j}) 
\ds - \kappa_q  U_q({\bf j}) U_p({\bfkappa} -{\bf j}) ) 
%\ d {\bf j}
\end{array}
\label{NS_fourier}
\eea
All the components ${\bf U}({\bfkappa}) \approx  \partial {\bf U}({\bfkappa})/ {\partial t} /\Lambda$ are random variables distributed according to certain distribution functions, which are statistically orthogonal each other \cite{Batchelor53}.

Thanks to the local isotropy, $\bf u$ is sum of several dependent random variables which are identically distributed \cite{Batchelor53}, therefore $\bf u$ tends to a gaussian variable \cite{Lehmann99}, and ${\bf U}({\bfkappa})$ satisfies the Lindeberg condition, a very general necessary and sufficient condition for satisfying the central limit theorem \cite{Lehmann99}. 
This condition does not apply to the Fourier coefficients of 
$\Delta {\bf u}$. In fact, since $\Delta {\bf u}$ is the difference between two dependent gaussian variables, its PDF could be a non gaussian
distribution function.
In ${\bf x}=0$, the velocity difference $\Delta {\bf u} ({\bf r}) \equiv
(\Delta u_1, \Delta u_2, \Delta u_3)$ is given by
\bea
\Delta u_p \hspace{-1.mm} \approx \hspace{-1.mm}   \frac{1}{\Lambda} 
\sum_{\bfkappa} \frac{\partial  U_p ({\bfkappa})} {\partial t} 
({\mbox e}^{i {\bfkappa}\cdot {\bf r}} - 1)    \equiv L + B + P + N
\eea
This fluctuation consists of the contributions appearing into Eq. (\ref{NS_fourier}):
in particular, $L$ represents the sum of all linear terms due to the viscosity
and $B$ is the sum of all bilinear terms arising from inertia and pressure
 forces. $P$ and $N$  are, respectively, the sums of definite positive 
and negative square terms, which derive from inertia and pressure forces.
The quantity $L+B$ tends to a gaussian random variable being 
the sum of statistically orthogonal terms \cite{Madow40, Lehmann99}, while $P$ and $N$ do
 not, as they are linear combinations of squares \cite{Madow40}.
 Their general  expressions are  \cite{Madow40}
\bea
\begin{array}{l@{\hspace{+0.2cm}}l}
 P = P_0 + \eta_1  +  \eta_2^2   \\\\
 N = N_0 + \zeta_1 +  \zeta_2^2  
\end{array} 
\label{nn}
\eea
where $P_0$ and $N_0$ are constants, and $\eta_1$, $\eta_2$, $\zeta_1$ and  $\zeta_2$ are four different centered random gaussian variables. 
Therefore, the fluctuation $\Delta u_p$ with zero average reads as 
\bea
\begin{array}{l@{\hspace{+0.2cm}}l}
\ds \Delta {u}_p  = 
\psi_1({\bf r}) {\xi} + \psi_2({\bf r})  
\ds \left( \chi  ( {\eta}^2-1 )  -  ( {\zeta}^2-1 )  \right) 
\end{array}
\label{fluc3}
\eea
where $\xi$, ${\eta}$ and $\zeta$ are independent centered random variables
which have gaussian distribution functions with standard deviation equal to the unity.
The parameter $\chi$ is a function of Reynolds number, 
whereas $\psi_1$ and $\psi_2$ are functions of space
coordinates, which also depend on the Reynolds number. 

At the Kolmogorov scale the order of magnitude of the velocity fluctuations is ${u_K}^2 \tau/\ell$, with $\tau = 1/\Lambda$, and $\psi_2$ is negligible because is due to the inertia forces: this immediately identifies $\psi_1 \approx {u_K}^2 \tau/\ell$.
\\
On the contrary, at the Taylor scale, $\psi_1$ is negligible and the order of magnitude of the velocity fluctuations is $u^2 \tau/\lambda_T$, therefore $\psi_2 \approx u^2 \tau/\lambda_T$ and the ratio $\psi_2 / \psi_1$ is a function of $R_\lambda$
\bea
\psi({\bf r}, R_{\lambda}) = \frac{\psi_2 ({\bf r})}{\psi_1({\bf r})} 
\approx \frac{u^2 \ell}{{u_K}^2 \lambda_T} =  
\sqrt{\frac{R_{\lambda}}{15 \sqrt{15}}} \
\hat{\psi}({\bf r}, R_{\lambda})
\label{Rl}
\eea
where 
$
\ds \hat{\psi}({\bf r}, R_{\lambda}) = O(1)
\label{R2}
$, 
is a function which has to be determined.
Hence, the longitudinal velocity difference $\Delta {u}_r$, is written as
\bea
\begin{array}{l@{\hspace{+0.2cm}}l}
\ds \frac {\Delta {u}_r}{\sqrt{\langle {\Delta {u}_r}^2} \rangle} =
\ds \frac{   {\xi} + \psi \left( \chi ( {\eta}^2-1 )  -  
\ds  ( {\zeta}^2-1 )  \right) }
{\sqrt{1+2  \psi^2 \left( 1+ \chi^2 \right)} } 
\end{array}
\label{fluc4}
\eea 
The quadratic term at the right hand side of Eq. (\ref{fluc4}) represents
the velocity fluctuations at the bigger scales, and there is no physical reason 
for which this must be bounded between same limits. Consequentely, 
$\chi$ must be a definite positive function of $R_\lambda$. 

Equation (\ref{fluc4}) gives the mathematical structure of 
$\Delta {u}_r$, whose dimensionless statistical moments 
are easily calculated considering that $\xi$, $\eta$ and 
$\zeta$ are independent gaussian variables
%These moments are 
%expressed in terms of $\chi$ and $\psi$
\bea
\begin{array}{l@{\hspace{+0.2cm}}l}
\ds H_n \equiv \frac{\left\langle \Delta u_r^n \right\rangle}
{\left\langle \Delta  u_r^2\right\rangle^{n/2} }
= 
\ds \frac{1} {(1+2  \psi^2 \left( 1+ \chi^2 \right))^{n/2}} \\\\
\ds \sum_{k=0}^n 
 \binom{n}{k} \psi^k
 \langle \xi^{n-k} \rangle 
  \langle (\chi(\eta^2 -1) - (\zeta^2 -1 ) )^k \rangle 
\end{array}
\label{m1}
\eea
where
\bea
\begin{array}{l@{\hspace{+0.2cm}}l}
\ds   \langle (\chi(\eta^2 -1) - (\zeta^2 -1 ) )^k \rangle = \\\\
\ds \sum_{i=0}^k 
 \binom{k}{i} (-\chi)^i 
 \langle (\zeta^2 -1 )^i \rangle 
 \langle (\eta^2 -1 )^{k-i} \rangle \\\\
\ds  \langle (\eta^2 -1 )^{i} \rangle = 
\sum_{l=0}^i 
\binom{i}{l} (-1)^{l}
\langle \eta^{2(i-l)} \rangle 
 \end{array}
\label{m2}
\eea
In particular, the third moment or skewness, $H_3$,
which is responsible for the energy cascade, is
\bea
\ds H_3= \frac{  8  \psi^3 \left( \chi^3 - 1 \right) }
 {\left( 1+2  \psi^2 \left( 1+ \chi^2 \right) \right)^{3/2}  }
\label{H_3}
\eea 
For $\chi \ne$ 1, the skewness and all the odd order moments are different from zero,
and for $n>3$, all the absolute moments are rising functions of $R_{\lambda}$, 
thus $\Delta u_r$ exhibits an intermittency whose entity increases 
with the Reynolds number. 
If $H_3$ and $\chi$ were both known, the other statistical
moments can be consequentely calculated with Eq. (\ref{m1}).
\begin{figure}
\vspace{-0.mm}
\centering
\includegraphics[width=0.45\textwidth]{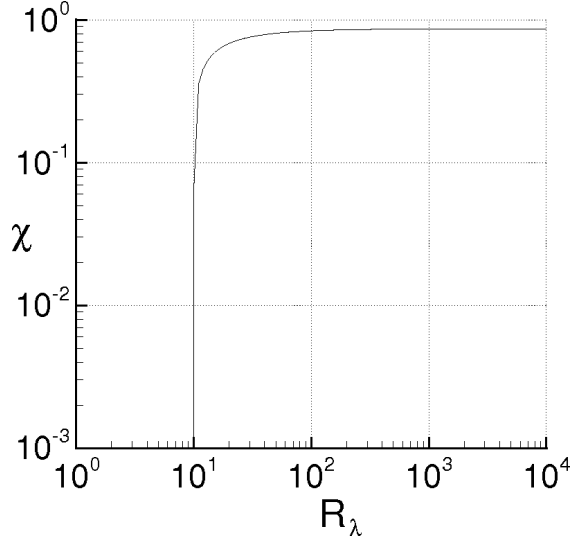}
\vspace{-2.mm}
\caption{Parameter $\chi$ plotted as the function of $R_{\lambda}$.}
\vspace{-0.mm}
\label{figura_2}
\end{figure}
The function $\psi (r, R_\lambda)$ is determined for $\Delta u_r$ 
from Eqs. (\ref{H_3}) and (\ref{H_3_01}). 
For $r$=0, one obtains the relationship
\bea
\frac{  8  {\psi_0}^3 \left(  1-\chi^3 \right) }
 {\left( 1+2  {\psi_0}^2 \left( 1+ \chi^2 \right) \right)^{3/2}  }
= \frac{3}{7}
\label{sk1}
\eea
$\psi_0 = \psi(0, R_{\lambda})=$O(1), is given by Eq. (\ref{Rl}), where, its exact value
 has to be calculated, whereas $\chi$ is a positive function of $R_{\lambda}$
 which must also be  determined. 
 To determine such quantities, note that Eq.(\ref{sk1}) is an algebraic
 relationship which gives $\chi$ in terms of $R_{\lambda}$, as shown 
in  Fig. \ref{figura_2}.
In any case, $\chi$ exhibits the limit $\chi \simeq$ 0.86592  for 
$R_{\lambda} \rightarrow \infty$, whereas $R_{\lambda}$ admits the minimim  $(R_\lambda)_{min}$ which depends on $\hat{\psi}_0$. Below such minimum, Eq. (\ref{sk1}) does not admit solutions with $\chi >0$. Then, according to the analysis of section \ref{s3}, $\hat{\psi}_0$ is chosen in such a way that $(R_\lambda)_{min}$ = 10.12 as shown in 
Fig. \ref{figura_2}, resulting $\hat{\psi}_0 \simeq 1.075$.
Now, all the moments of $\Delta u_r$ can be calculated 
by Eqs. (\ref{m1}) and (\ref{m2}) in terms of $R_{\lambda}$.

The PDF of $\Delta u_r$ is expressed through the Frobenious-Perron equation
\bea
\begin{array}{l@{\hspace{+0.3cm}}l}
F(\Delta {u'}_r) = \hspace{-1.mm}
\ds \int_\xi \hspace{-1.mm}
\int_\eta  \hspace{-1.mm}
\int_\zeta \hspace{-1.mm}
p(\xi) p(\eta) p(\zeta) \
\delta \left( \Delta u_r\hspace{-1.mm}-\hspace{-1.mm}\Delta {u'}_r \right)   
d \xi d \eta d \zeta
\end{array}
\label{frobenious_perron}
\eea 
where $\Delta {u}_r$ is calculated with Eq. (\ref{fluc4}), $\delta$ is the Dirac delta and $p$ is a gaussian PDF whose average value and standard deviation are equal to 0 and 1, respectively.

\bigskip

For non-isotropic turbulence or in more complex cases
with boundary conditions, the velocity spectrum could not satisfy the
Lindeberg condition, thus the velocity will be not distrubuted following 
a Gaussian PDF, and Eq. (\ref{fluc3}) changes its analytical form and can incorporate more intermittant terms \cite{Lehmann99} which give the deviation with respect to the isotropic turbulence.
Hence, the absolute statistical moments of $\Delta {u}_r$ will be greater than those calculated with Eq. (\ref{fluc4}), indicating that, in a more complex situation than the isotropic turbulence, the intermittancy of $\Delta {u}_r$ can be significantly stronger.

\section{Results and discussion  \label{s8}}

The results calculated with the proposed theory are now presented.

As the first result, the evolution in time of the correlation function is calculated with the proposed closure of the von K\'arm\'an-Howarth equation (Eq. (\ref{vk6})), where the boundary conditions are given by Eq. (\ref{bc}).
The turbulent kinetic energy and the spectrums $E(\kappa)$ and $T(\kappa)$ are calculated
with Eq. (\ref{ke}) and Eqs. (\ref{Ek}), respectively.
The calculation is carried out for the initial Reynolds number of 
$Re = u(0) L_r/ \nu$ = 2000, where $L_r$ and $u(0)$ are, respectively, 
the characteristic dimension of the problem and the initial velocity standard deviation. The initial condition is a gaussian correlation function with 
$\lambda_T/L_r$ = $1/(2 \sqrt{2})$. 
The dimensionless time of the problem is defined as $\bar{t} = t \ u(0)/L_r$.

Equation (\ref{vk}) was numerically solved adopting the Crank-Nicholson integrator scheme with variable time step, where the discretization of the space domain is made  by $N-1$  intervals of the same amplitude $\Delta r$. 
This corresponds to a discretization of the Fourier space made by $N-1$ subsets in the interval $\left[ 0, \kappa_M \right]$, where $\kappa_M$ = $\pi/(2 \Delta r)$.
For the adopted initial Reynolds number, the choice $N$ = 1500, gives an adequate discretization, which provides $\Delta r < \ell$, for the whole simulation. 
During the simulation, $T(\kappa)$ must identically satisfy Eq.(\ref{tk0})
(see Appendix) which states that $T(\kappa)$ does not modify the kinetic energy.
To verify Eq.(\ref{tk0}), the integral of $T(\kappa)$ 
is calculated with the trapezes rule from $0$ until to $\kappa_M$, at each time step,
therefore, the simulation will be considered to be accurate as long as
\bea
\int_0^{\kappa_M} T(\kappa) d \kappa \simeq \int_0^{\infty}
 T(\kappa) d \kappa = 0
\label{tk0a}
\eea   
namely, when the energy is distributed for $\kappa < \kappa_M$.
As the simulation advances, according to Eq. (\ref{vk6}),
the energy cascade determines variations of $E(\kappa)$ and $T(\kappa)$ at the higher wave-numbers, then Eq. (\ref{tk0a}) will hold until to a
certain time. For this reason, the simulation is stopped as soon as the
following condition is achieved \cite{NAG}
\bea
\ds \vert \int_0^{\kappa_M} T(\kappa) d \kappa \vert >
 \frac{1}{N^2} \int_0^{\kappa_M} \vert T(\kappa) \vert  d \kappa 
\eea
At the end of several simulations, we have $\Delta r \approx 0.8 \ \ell$,
and, in this situation, the energy spectrum is here considered to be fully developed.
\begin{figure}
\vspace{-0.mm}
	\centering
\includegraphics[width=0.45\textwidth]{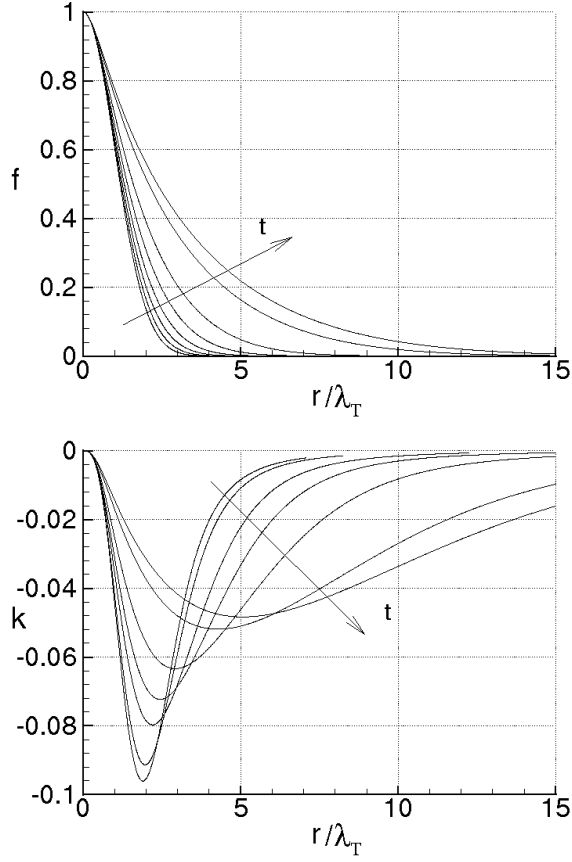}
\vspace{-0.mm}
\caption{Correlation functions, $f$ and $k$ versus the separation distance at the times of simulation $\bar{t}$ = 0, 0.1, 0.2, 0.3, 0.4, 0.5, 0.6, 0.63.}
\vspace{-0.mm}
\label{figura_6}
% f_k.jpg: 100dpi, width=7.62cm, height=10.80cm, bb=0 0 300 425
\end{figure}

The diagrams of Fig. \ref{figura_6} show the correlation functions $f(r)$ and $k(r)$ vs. the dimensionless distance $r/\lambda_T$, at different times of simulation. 
The kinetic energy and Taylor scale diminish according to Eqs. (\ref{vk6}) and (\ref{ke}), thus $f(r)$ and $k(r)$ change in such a way that the length scales 
associated to their variations diminish as the time increases, whereas the maximum of 
$\vert k \vert$ decreases. 
At the final instants of the simulation, one obtains that 
$f - 1 =$ O( $r^{2/3}$) for $r/\lambda_T=$ O(1), whereas the maximum of 
$\vert k \vert$ is about 0.05. 
These results are in very good agreement with the numerous data of the literature \cite{Batchelor53} which concern the evolution of correlation function and energy spectrum.
\begin{figure}
\vspace{-0.mm}
	\centering
\hspace{-0.mm}
\includegraphics[width=0.49\textwidth]{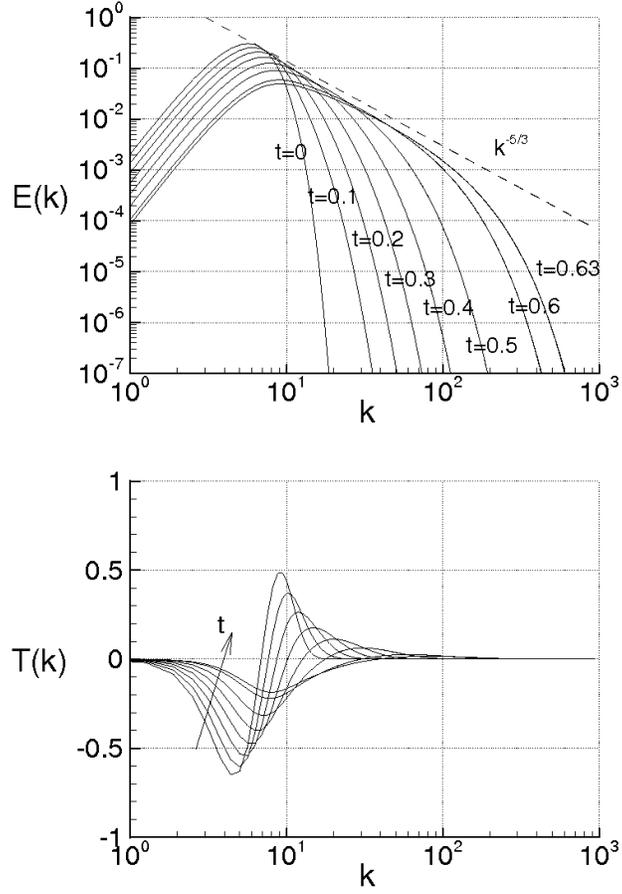}
\vspace{-0.mm}
\caption{Plot of $E(\kappa)$ and $T(\kappa)$ at the diverse times of simulation.}
% Ek_Tk.jpg: 100dpi, width=10.16cm, height=14.40cm, bb=0 0 400 567
\vspace{-0.mm}
\label{figura_7}
\end{figure}
Figure \ref{figura_7} shows the diagrams of $E(\kappa)$ and $T(\kappa)$ 
for the same times, where the dashed line in the plot of $E(\kappa)$, 
represents the $-5/3$ Kolmogorov law \cite{Kolmogorov41}. 
\begin{figure}
\vspace{+0.mm}
	\centering
\hspace{-0.mm}
\includegraphics[width=0.42\textwidth]{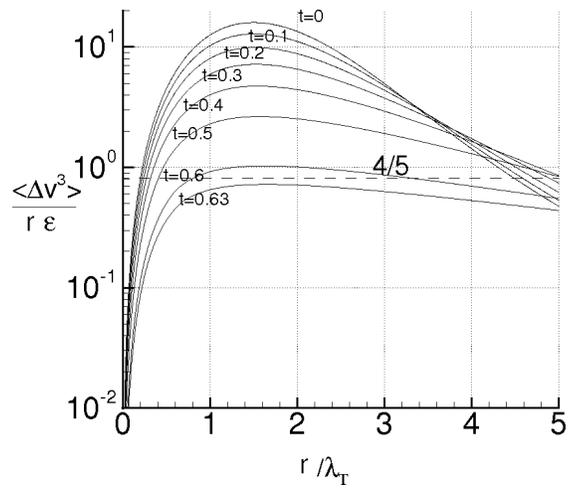}
\vspace{-0.mm}
\caption{The Kolmogorov function versus $r/\lambda_T$ for different times of simulation. The dashed line indicates  the value  4/5.}
\vspace{-0.mm}
% f_kolm.jpg: 100dpi, width=10.16cm, height=7.16cm, bb=0 0 400 282
\label{figura_8}
\end{figure}
The spectrums $E(\kappa)$ and $T(\kappa)$ vary according to Eqs. (\ref{vk6}) and (\ref{Ek}), and, at the end of simulation, $E(\kappa)$ is about parallel to the dashed line in an opportune interval of the wave-numbers which defines the so called 
inertial range of Kolmogorov.
This arises from the developed correlation function, which behaves like
$f -1$ = O ($r^{2/3}$) for $r = O(\lambda_T)$.
\begin{figure}
\vspace{-0.mm}
	\centering
\hspace{+10.mm}
\includegraphics[width=0.45\textwidth]{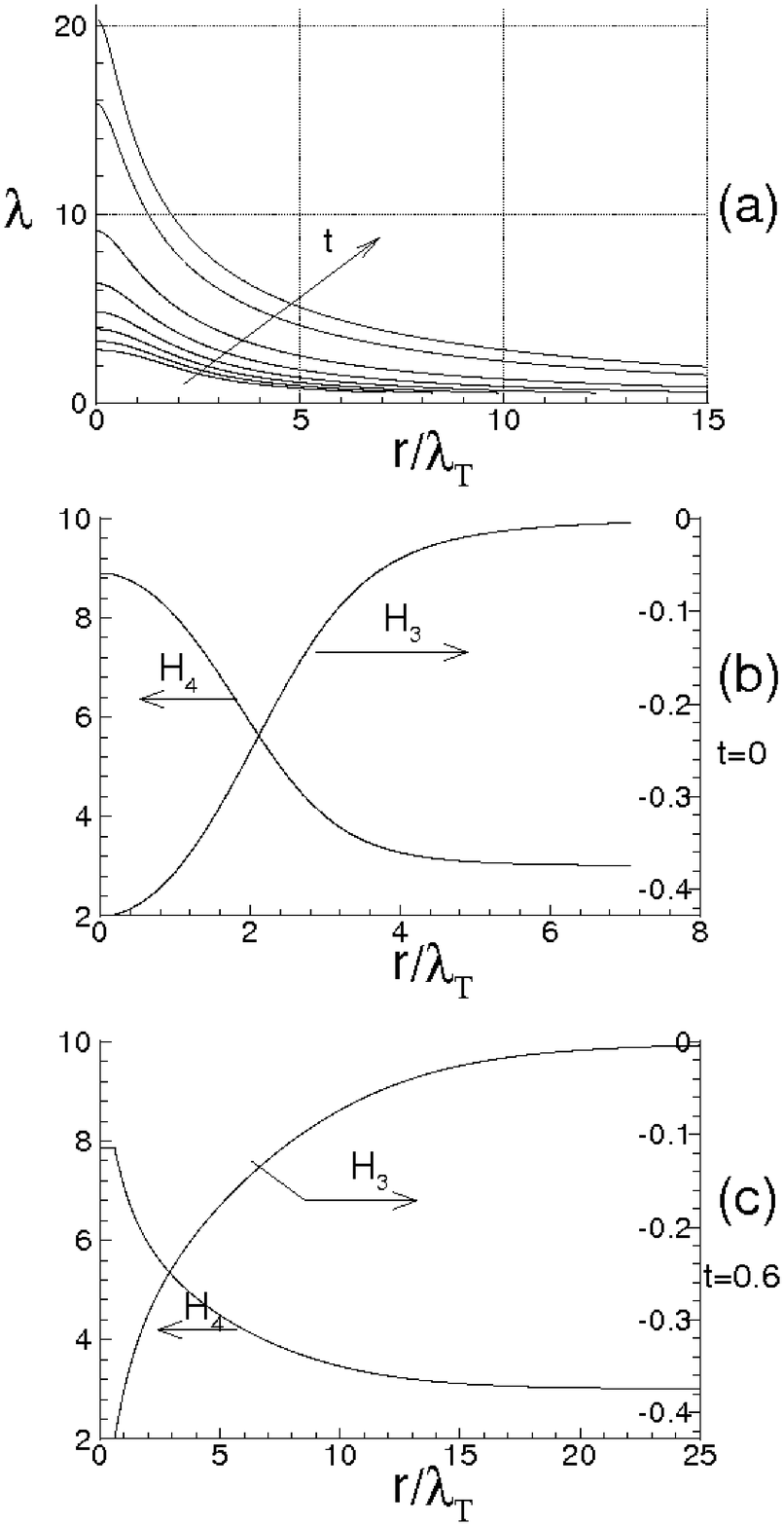}
% lambda_H.jpg: 100dpi, width=10.16cm, height=14.40cm, bb=0 0 400 567
\vspace{-0.mm}
\caption{(a) Maximum finite size Lyapunov exponent at the times of simulation $\bar{t}$ = 0, 0.1, 0.2, 0.3, 0.4, 0.5, 0.6, 0.63; (b) and (c) skewness and Flatness versus $r/\lambda_T$ at t = 0 and t = 0.6, respectively.}
\vspace{-0.mm}
\label{figura_9}
\end{figure}

Next, the Kolmogorov function $Q(r)$ and Kolmogorov constant $C$, are determined with
the proposed theory, using the previous results of the simulation. 

Following the Kolmogorov theory, the Kolmogorov function, which is defined as  
\bea 
\ds Q(r) = - \frac{\langle {\Delta u_r}^3 \rangle} { r \varepsilon}
\label{k_f}
\eea
is constant with respect to $r$, and is equal to 4/5 as long as $r/\lambda_T = O(1)$. 
As shown in Fig. \ref{figura_8}, for $\bar{t} =0$, the maximum of $Q(r)$ is much greater than 4/5 and its variations with $r/\lambda_T$ can not be neglected.
This is due to the arbitrary choice of the initial correlation function.
At the successive times, the variations of $f$ determine that the maximum of $Q(r)$ and its variations decrease until to the final instants, where, with the exception of $r/\lambda_T \approx 0$, $Q(r)$ exhibits a qualitatively flat shape in a wide range of $r/\lambda_T$, with a maximum which is quite close to 0.8. 

The Kolmogorov constant $C$ is also calculated by definition
\bea
E(\kappa) = C \frac{\varepsilon^{2/3} } {\kappa^{5/3}}
\label{k_c}
\eea 
This is here determined, as the value of $C$ which makes the curve represented by  
Eq. (\ref{k_c}) to be tangent to the energy spectrum $E(\kappa)$ 
previously calculated.
At end simulation, $C \simeq $ 1.932, namely $C$ and $Q_{max}$ agree very well to the corresponding quantities known from the literature.
For the same simulation, Fig. \ref{figura_9}a shows the maximal finite scale Lyapunov exponent, calculated with Eq. (\ref{lC}), where $\lambda$ varies according to $f$.
For $t = 0$, the variations of $\lambda$ are relatively small because of the 
adopted initial correlation function which is a gaussian, 
whereas as the time increases, the variations of $f$ determine sizable increments of 
$\lambda$ and of its slope in proximity of the origin. 
Then, for developed spectrum, since $f -1$ = O($r^{2/3}$), 
the maximal finite scale Lyapunov exponent behaves like $\lambda \approx r^{-2/3}$.
Thus, the diffusivity coefficient associated to the relative motion between two fluid particles,  defined as $D(r) \propto \lambda r^2$, here satisfies the famous
Richardson scaling law $D(r) \approx r^{4/3}$\cite{Richardson26}.
\begin{figure}
\vspace{-0.mm}
	\centering
\includegraphics[width=0.48\textwidth]{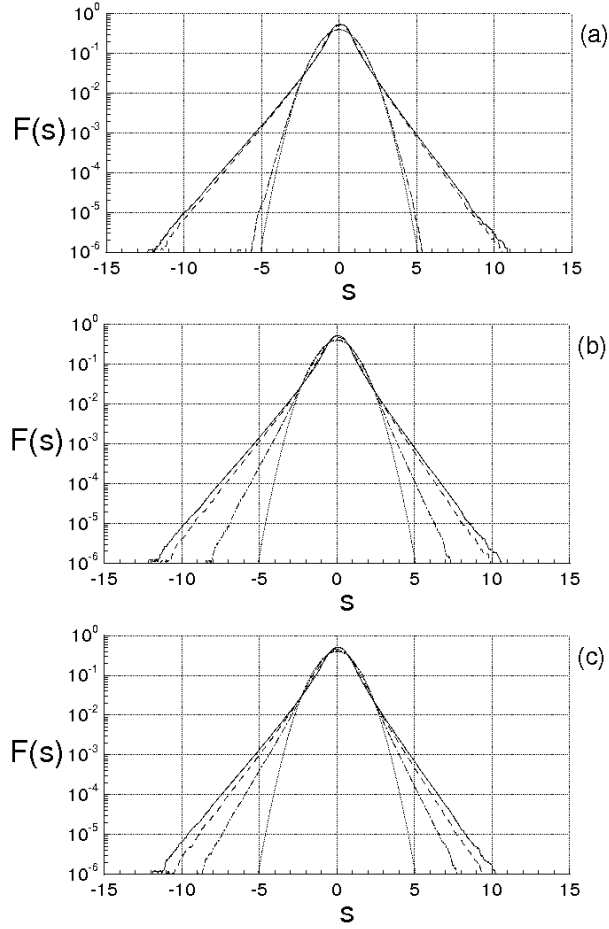}
% pdf_dv_abc.jpg: 100dpi, width=11.43cm, height=16.18cm, bb=0 0 450 637
\vspace{-0.mm}
\caption{PDF of the velocity difference fluctuations at the times $\bar{t}=$0 (a), $\bar{t}=$ 0.5 (b) and $\bar{t}=$0.6 (c). 
Continuous lines are for $r=$0, dashed lines are for $r/\lambda_T =$1, 
dot-dashed lines are for  $r/\lambda_T =$5, dotted lines are for gaussian PDF.}
\vspace{-0.mm}
\label{figura_10}
\end{figure}

In the diagrams of Figs. \ref{figura_9}b and \ref{figura_9}c, skewness and flatness of $\Delta u_r$ are shown in terms of $r$ for $\bar{t}$ = 0 and 0.6.
The skewness, $H_3$ is first calculated with Eq. (\ref{H_3_01}), 
then $H_4$ has been determined using Eq. (\ref{m1}). 
At $\bar{t}=0$, $\vert H_3 \vert$ starts from 3/7 at the origin with small slope, then decreases until to reach small values. $H_4$ also exhibits small derivatives near the origin, where $H_4\gg$ 3, thereafter it decreases more rapidly than $\vert H_3 \vert$. 
At $\bar{t} = $0.6, the diagram importantly changes and exhibits different shapes.
The Taylor scale and the corresponding Reynolds number are both diminished, so that the variations of $H_3$ and $H_4$ are associated to smaller distances, whereas the 
flatness at the origin is slightly less than that at $t =0$. Nevertheless, these variations correspond to higher $r/\lambda_T$ than those for $t$ = 0, and also in this case, $H_4$ reaches the value of 3 more rapidly than $H_3$ tends to zero.

The PDFs of $\Delta u_r$ are calculated with Eqs. (\ref{frobenious_perron})
and (\ref{fluc4}), and are shown in Fig. \ref{figura_10} in terms of the dimensionless 
abscissa 
\bea
\ds s = \frac{\Delta u_r} 
{ \langle \Delta u_r^2 \rangle^{1/2}  }
\nonumber
\eea
where, these distribution functions are normalized, in order that their standard 
deviations are equal to the unity.
The figure  represents the distribution functions of $s$ for several
$r/\lambda_T$, at $\bar{t}$ = 0, 0.5 and 0.6, where the dotted curves represent the gaussian distribution functions.
The calculation of $H_3(r)$ is first carried out with Eq. (\ref{H_3_01}), then the function $\psi(r, R_\lambda)$ is identified through Eq. (\ref{H_3}), and finally the PDF is obtained with  Eq. (\ref{frobenious_perron}).
For $t$ = 0 (see Fig. \ref{figura_10}a) and according to the evolutions of $H_3$ and $H_4$, the PDFs calculated  at $r/\lambda_T=$ 0 and 1, are quite similar each other, whereas for $r/\lambda_T=$ 5, the PDF is an almost gaussian function.
Toward the end of the simulation, (see Fig. \ref{figura_10}b and c), the two PDFs
calculated at $r/\lambda_T=$ 0 and 1, exhibit more sizable differences, whereas for $r/\lambda_T=$ 5, the PDF differs very much from a gaussian PDF. 
This is in line with the plots of $H_3(r)$ and $H_4(r)$ of Fig. \ref{figura_9}.
\begin{figure}
\vspace{-0.mm}
	\centering
\hspace{-0.mm}
\includegraphics[width=0.45\textwidth]{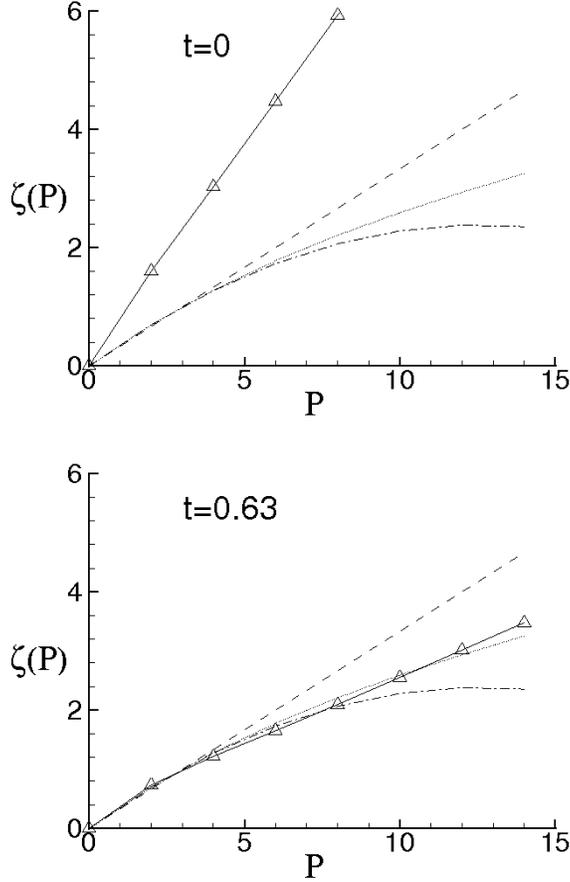}
% zp.jpg: 100dpi, width=7.62cm, height=10.80cm, bb=0 0 300 425
\vspace{-0.mm}
\caption{Scaling exponents of longitudinal velocity difference versus the order moment at different times. Continuous lines with solid symbols are for the present data. Dashed lines are for Kolmogorov K41 data \cite{Kolmogorov41}. Dashdotted lines are for Kolmogorov K62 data \cite{Kolmogorov62}.
 Dotted lines are for She-Leveque data \cite{She-Leveque94}}
\vspace{-0.mm}
\label{figura_11}
\end{figure}

\bigskip

Next, the spatial structure of $\Delta u_r$, given by Eq. (\ref{fluc4}), is analyzed 
using the previous results of the simulation. 
According to the various works \cite{Kolmogorov62, She-Leveque94, Benzi91}, $\Delta u_r$ behaves quite similarly to a multifractal system, where $\Delta u_r$ obeys to a law of the kind 
$
\Delta u_r(r) \approx r^q
$
where the exponent $q$ is a fluctuating function of space.
This implies that the statistical moments of $\Delta u_r(r)$ are expressed through 
different scaling exponents $\zeta(P)$ whose values depend on the moment order $P$, i.e.
\bea
\left\langle \Delta u_r^{P}(r) \right\rangle  = A r^{\zeta(P)}
\label{fractal}
\eea
These scaling exponents are here identified through a best fitting procedure, 
in the interval $2 \ell < r < \lambda_T$, where the statistical moments of   
$\Delta u_r(r)$ are calculated with Eqs. (\ref{m1}). 
Figure \ref{figura_11} shows the comparison between the scaling exponents here obtained 
(continuous lines with solid symbols) and those of the Kolmogorov theories K41 \cite{Kolmogorov41} (dashed lines) and K62 \cite{Kolmogorov62} (dashdotted lines), and those given by She-Leveque \cite{She-Leveque94} (dotted curves).
At $t=$ 0, the slope of $\zeta(P)$ is about constant, whereas the values of $\zeta(P)$ are very different from those calculated by the various authors. This means that, for the chosen initial correlation function, $\Delta u_r(r)$ behaves like a simple fractal system, where $\zeta(P) \propto P$. Again, this result 
depends on the fact that, at the initial times, the energy spectrum is not developed.
As the time increases, the correlation function changes 
causing variations in the statistical moments of $\Delta u_r(r)$.
As result, $\zeta(P)$ gradually diminish and exhibit a 
variable slope which depends on the moment order $P$, until to reach the situation of Fig. \ref{figura_11}b, where the simulation is just ended. 
The correlation function and the dimensionless moments of $\Delta u_r(r)$ are changed, thus the plot of $\zeta(P)$ shows that near the origin, $\zeta(P) \simeq P/3$, whereas elsewhere the values of $\zeta(P)$ are in agreement with the She-Leveque results, confirming that $\Delta u_r(r)$ behaves like a multifractal system.

Other simulations with different initial correlation functions and Reynolds
numbers have been performed, and all of them lead to analogous results,
in the sense that, at the end of the simulations, the diverse quantities 
such as $Q(r)$, $C$ and $\zeta(P)$ are quite similar to those just calculated.
For what concerns the effect of the Reynolds number, its increment determines
a wider Kolmogorov inertial range and a smaller dissipation energy rate in accordance
to Eq. (\ref{ke}), whereas the shapes of the various energy spectrums
remain qualitatively unaltered with respect to Fig. \ref{figura_7}.

\bigskip
\begin{table}[b] 
  \begin{center} 
  \begin{tabular}{lrrrr} 
Moment \ & $R_\lambda \approx 10$ \ & $R_\lambda=10^2$ \ & $R_\lambda=10^3$ \ & Gaussian\\[2pt] 
Order    &  P. R.                 & P. R.            &  P. R. \           & Moment \\[2pt] 
3        & -.428571               & -.428571         & -.428571          & 0      \\
4        &   3.96973              &  7.69530         & 8.95525           & 3      \\
5        & -7.21043               &  -11.7922        & -12.7656          & 0      \\
6        &  42.4092               &  173.992         & 228.486           & 15     \\
7        & -170.850               &  -551.972        & -667.237          & 0      \\
8        &  1035.22               &  7968.33         & 11648.2           & 105    \\
9        &  -6329.64              &  -41477.9        & -56151.4          & 0      \\
10       & 45632.5                &  617583.         & 997938.           & 945    \\
 \end{tabular}
\caption{Dimensionless statistical moments of $F(\partial u_r/\partial r)$ at different
Taylor scale Reynolds numbers. P.R. as for ''present results''.}
  \end{center} 
  \vspace{-5.mm}
\end{table} 

\begin{figure}[t]
\vspace{-0.mm}
	\centering
\includegraphics[width=0.45\textwidth]{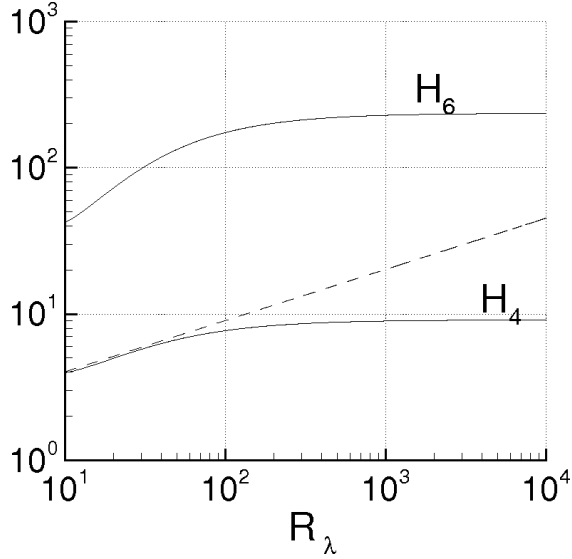}
\vspace{-0.mm}
\caption{Dimensionless moments $H_4(0)$ and $H_6(0)$ plotted vs. $R_{\lambda}$.
Continuous lines are for the present results. 
The dashed line is the tangent to the curve of $H_4(0)$ in $R_\lambda \approx$ 10.}
\vspace{-0.mm}
\label{figura_3}
% H_Re.jpg: 100dpi, width=8.89cm, height=6.27cm, bb=0 0 350 247
\end{figure}

In order to study the evolution of the intermittancy vs. the Reynolds number,
Table 1 gives the first ten statistical moments of $F(\partial u_r/\partial r)$. 
These are calculated with Eqs. (\ref{m1}) and (\ref{m2}), for $R_{\lambda}$ = 10.12, 100 and 1000, and are shown in comparison with those of a gaussian distribution function.
It is apparent that a constant nonzero skewness of the longitudinal velocity derivative, causes an intermittancy which rises with $R_\lambda$ 
(see Eq. (\ref{fluc4})).
More specifically, Fig. \ref{figura_3} shows the variations of $H_4(0)$ and $H_6(0)$
(continuous lines) in terms of $R_\lambda$, calculated with Eqs. (\ref{m1}) 
and (\ref{m2}), with $H_3(0) = -3/7$.
These moments are rising functions of $R_{\lambda}$ for 
10 $\lesssim R_{\lambda} \lesssim$ 700, whereas for higher $R_{\lambda}$ these tend to
the saturation and such behavior also happens for the other absolute moments.
According to Eq. (\ref{m1}), in the interval 10 $ \lesssim R_{\lambda} \lesssim$  70, 
$H_4$ and $H_6$ result to be about proportional to $R_{\lambda}^{0.34}$ and $R_{\lambda}^{0.78}$,
respectively, and the intermittancy increases with the Reynolds number until to 
$R_{\lambda} \approx$ 700,  where it ceases to rise so quickly. 
\begin{figure}[t]
\vspace{-0.mm}
	\centering
\includegraphics[width=0.48\textwidth]{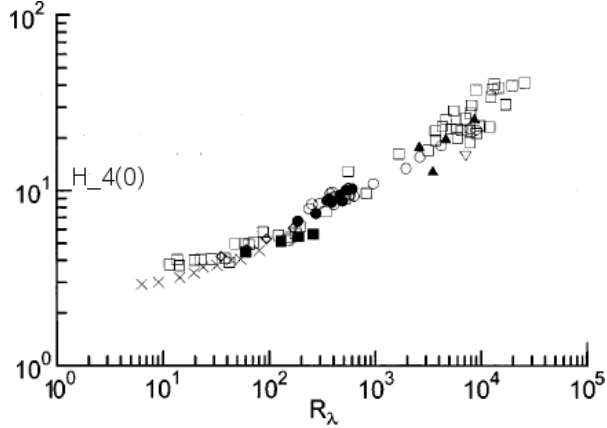}
\vspace{-0.mm}
\caption{Flatness $H_4(0)$ vs. $R_{\lambda}$. These data are from  Ref.\cite{Antonia97}.}
\vspace{-0.mm}
\label{antonia}
% H_Re.jpg: 100dpi, width=8.89cm, height=6.27cm, bb=0 0 350 247
\end{figure}
\begin{figure}[t]
\vspace{-0.mm}
	\centering
\includegraphics[width=0.50\textwidth]{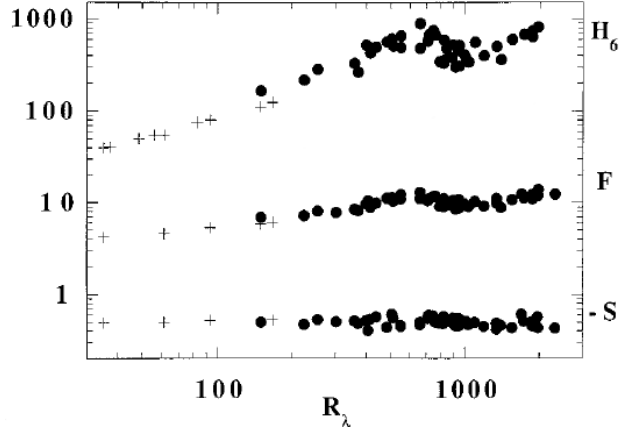}
\vspace{-0.mm}
\caption{Skewness $S = H_3(0)$, Flatness $F =H_4(0)$ and hyperflatness $H_6(0)$ vs. $R_{\lambda}$. These data are from  Ref.\cite{Tabeling97}.}
\vspace{-0.mm}
\label{tabeling0}
% H_Re.jpg: 100dpi, width=8.89cm, height=6.27cm, bb=0 0 350 247
\end{figure}
This behavior, represented by the continuous lines, depends on the fact that 
$\psi \approx \sqrt{R_\lambda}$, and results to be in very good agreement with the data 
of Pullin and Saffman \cite{Pullin93}, for 10 $\lesssim R_{\lambda} \lesssim$  100.
\begin{figure}[t]
\vspace{+0.mm}
\centering
\hspace{0.mm}
\includegraphics[width=0.45\textwidth]{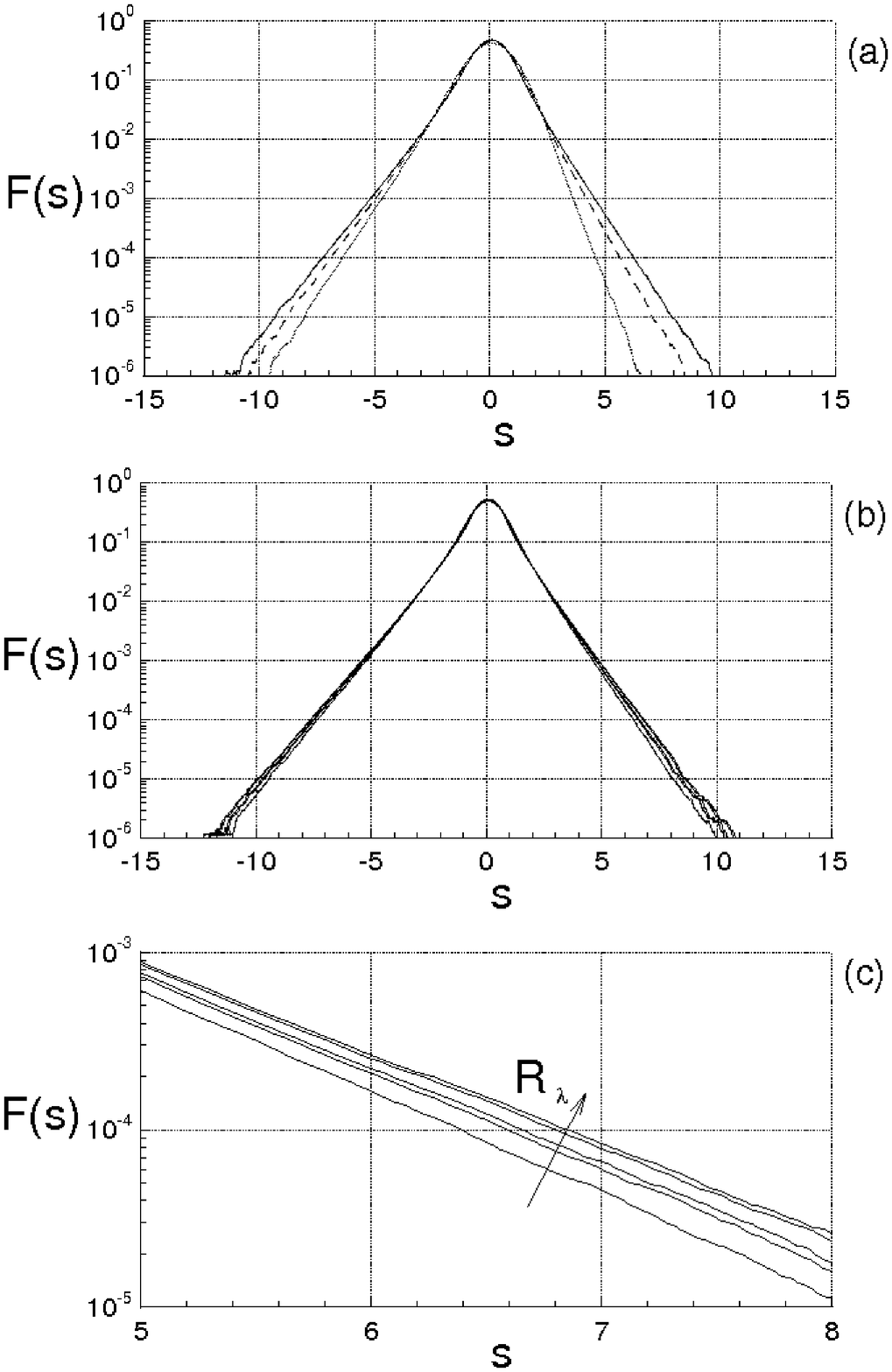}
\vspace{-0.mm}
\caption{Log linear plot of the PDF of $\partial u_r/\partial r$ for different
$R_\lambda$. (a): dotted, dashdotted and continuous lines are
for $R_\lambda$ = 15, 30 and 60, respectively. (b) and (c) PDFs for 
$R_{\lambda}$ = 255, 416, 514, 1035 and 1553. (c) represents an enlarged part
of the diagram (b) }
\vspace{-0.mm}
\label{figura_4}
% pdf_abc.jpg: 100dpi, width=10.16cm, height=14.40cm, bb=0 0 400 567
\end{figure}
\begin{figure}[t]
\vspace{-0.mm}
\centering
\hspace{0.mm}
\includegraphics[width=0.42\textwidth]{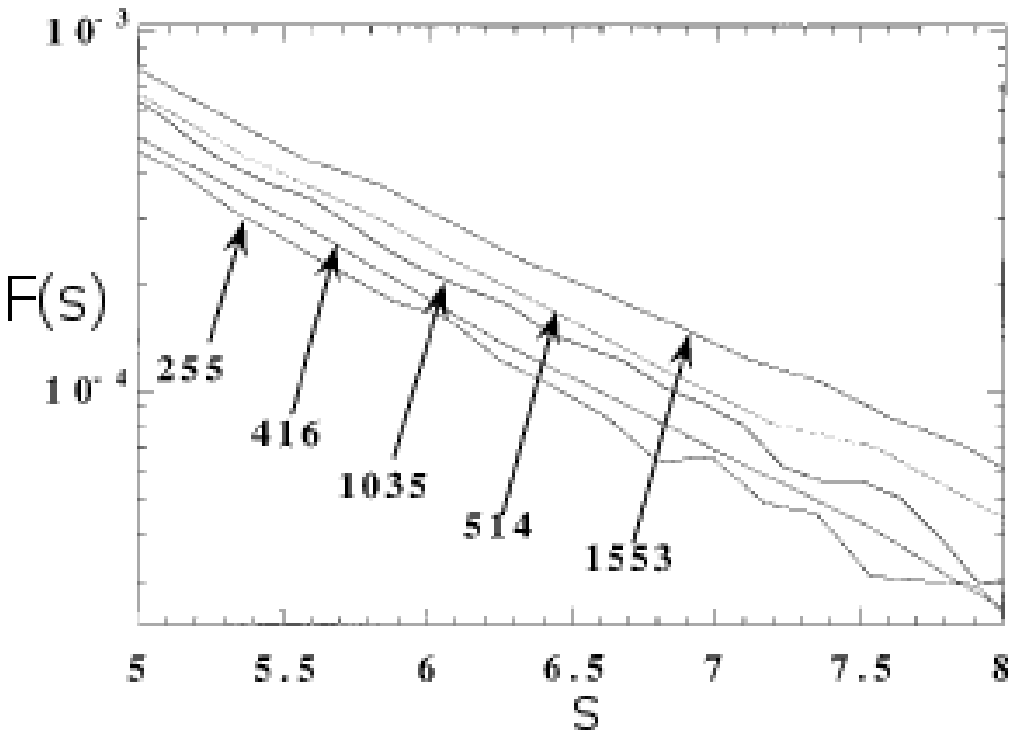}
\vspace{-0.mm}
\caption{PDF of $\partial u_r/\partial r$ for 
$R_{\lambda}$ = 255, 416, 514, 1035 and 1553. These data are from Ref. \cite{Tabeling97} }
\vspace{-0.mm}
\label{tabeling}
% pdf_abc.jpg: 100dpi, width=10.16cm, height=14.40cm, bb=0 0 400 567
\end{figure}
Figure \ref{figura_3} can be compared with the data collected by Sreenivasan and Antonia \cite{Antonia97}, which are here reported into Fig. \ref{antonia}.
These latter are referred to several measurements and simulations obtained in different situations which can be very far from the isotropy and homogeneity conditions. 
Nevertheless a comparison between the present results and those of 
Ref. \cite{Antonia97} is an opportunity to state if the two data exhibit 
elements in common.
According to  Ref. \cite{Antonia97}, the flatness monotonically rises with $R_\lambda$ with a rising rate which agrees with Eq. (\ref{m2}) for 
$10 \lesssim R_\lambda \lesssim 60$ (dashed line, Fig. \ref{figura_3}), whereas the skewness seems to exhibit minor variations.
Thereafter, $H_4$ continues to rise with about the same rate, 
without the saturation observed in Fig. \ref{figura_3}.
The weaker intermittancy calculated with the present theory arise from the isotropy which makes the velocity fluctuation a gaussian random variable, while,
as seen in sec. \ref{s4}, without the isotropy condition, the flatness of velocity 
and of velocity difference can be much greater than that of the isotropic case.

Again, the obtained results are compared with the data of Tabeling {\it et al} 
\cite{Tabeling96,  Tabeling97}, where, in an experiment using low temperature helium gas  between two counter-rotating cylinders (closed cell), the authors measure the PDF of $\partial u_r/\partial r$ and its moments.
Also in this case the flow can be quite far from to the isotropy condition.
In fact, these experiments pertain wall-bounded flows, where the walls could importantly influence the fluid velocity in proximity of the probe. 
The authors found that the higher moments than the third order, first increase with
$R_{\lambda}$ until to $R_{\lambda} \approx$ 700, then exhibit a lightly non-monotonic evolution with respect to $R_{\lambda}$, and finally cease their
 variations denoting a transition behavior (See Fig. \ref{tabeling0}). 
As far as the skewness is concerned, the authors observe small percentage variations. 
Although the isotropy does not describe the non-monotonic evolution near 
$R_{\lambda} =$ 700, the results obtained with Eq. (\ref{fluc4}) can be considered comparable with those of Refs. \cite{Tabeling96, Tabeling97}, resulting also
in this case, that the proposed theory gives a weaker intermittancy with respect to 
Refs. \cite{Tabeling96, Tabeling97}.

The normalized PDFs of $\partial u_r/\partial r$ are calculated with Eqs. (\ref{frobenious_perron}) and (\ref{fluc4}), and are shown in Fig. \ref{figura_4} 
in terms of the variable $s$, which is defined as  
\bea
\ds s = \frac{\partial u_r/\partial r} 
{ \left\langle (\partial u_r/\partial r)^2\right\rangle^{1/2}  }
\nonumber
\eea
Figure \ref{figura_4}a shows the diagrams for $R_{\lambda} =$ 15, 30 and 60, where
the PDFs vary in such a way that $H_3(0) = -3/7$.
\\
As well as in Ref. \cite{Tabeling97}, Figs. 4b and 4c give the PDF for
$R_{\lambda}$ = 255, 416, 514, 1035 and 1553,  where these last Reynolds numbers are calculated through the  Kolmogorov function given in Ref. \cite{Tabeling97}, with 
$H_3(0) = -3/7$.
In particular, Fig. \ref{figura_4}c represents the enlarged region of Fig. \ref{figura_4}b,  where the tails of PDF are shown for $5 < s < 8$.
According to Eq. (\ref{fluc4}), the tails of the PDF rise in the interval  
10 $\lesssim R_{\lambda} \lesssim$ 700, whereas at higher  $R_{\lambda}$, smaller variations occur.
Although the non-monotonic trend observed in Ref. \cite{Tabeling97}, 
Fig. \ref{figura_4}c shows that the values of the PDFs
calculated with the proposed theory, for $5 < s < 8$, exhibit the same
order of magnitude of those obtained by Tabeling {\it et al} 
\cite{Tabeling97} which are here shown in Fig. \ref{tabeling}.
\begin{figure}[t]
\vspace{0.mm}
 \centering
\hspace{-0.mm}
\includegraphics[width=0.50\textwidth]{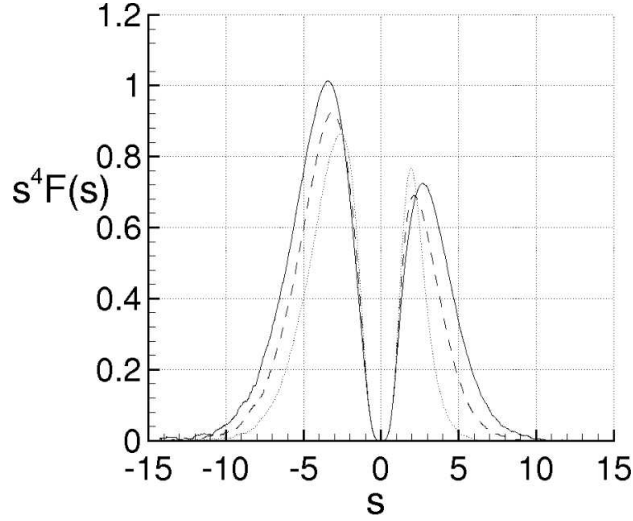}
\vspace{-0.mm}
\caption{Plot of the integrand $s^4 F(s)$ for different
$R_\lambda$. Dotted, dashdotted and continuous lines are
for $R_\lambda$ = 15, 30 and 60, respectively.}
\vspace{-0.mm}
\label{figura_5}
% dv4_f.jpg: 100dpi, width=10.16cm, height=7.16cm, bb=0 0 400 282
\end{figure}

Asymmetry and intermittency of the distribution functions are also 
represented through the integrand function of the $4^{th}$
order moment of PDF, which is
$
J_4(s) = s^4 F(s)
\nonumber
$
This function is shown in terms of $s$, in Fig. \ref{figura_5}, for
 $R_\lambda$ = 15, 30 and 60.

\section{\bf  Conclusions  \label{s9}}

The proposed theory is based on the Landau conjecture which states that  
the turbulence is caused by the bifurcations of the velocity field.

The obtained results confirm the capability of the proposed theory to
describe quite well the general properties of the turbulence. 
These results are here summarized:

\begin{enumerate}

\item The analysis of the bifurcations gives the connection
between number of bifurcations, length scales and Reynolds number
at the onset of the turbulence and allows to determine the minimum 
Taylor-scale Reynolds number for isotropic turbulence. 
This last one is about 10, and, below this value, the isotropic 
turbulence is not allowed.

\item The momentum equations written using the referential description allow 
the velocity fluctuation to be expressed by means of the
Lyapunov analysis of the kinematics of fluid deformation.

\item The Lyapunov analysis of the relative kinematics equations provides
an explanation of the physical mechanism of the energy cascade in turbulence. 
The non-ergodicity of  $d/dt ({\bf u}\cdot{\bf u}')$, due to the
fluid incompressibility, make possible that the inertia forces transfer the kinetic energy
between the length scales without changing the total kinetic energy.
This implies that the skewness of the longitudinal velocity derivative is a constant of the present theory and that the energy cascade mechanism does not depend on the Reynolds number.

\item The Fourier analysis of the velocity difference provides the statistics of
$\Delta u_r$. 
This is a non-Gaussian statistics, where the constant skewness of 
$\partial u_r/ \partial r$ implies that the other higher absolute moments increase with the Taylor-scale Reynolds number. 

\item The developed energy spectrums, calculated with the proposed closure of the von K\'arm\'an-Howarth equation, agrees quite well with the Kolmogorov law $\kappa^{-5/3}$ in a given interval of $\kappa$ which defines the inertial subrange of Kolmogorov. 

\item For developed energy spectrums, the Kolmogorov function is about constant in a wide range of separation distances and its maximum is quite close to 4/5, whereas the Kolmogorov constant is about equal to 1.93. 
As the consequence, the maximal finite scale Lyapunov exponent and the diffusivity coefficient,  vary according to the Richardson law when the separation distance is of the order of the Taylor scale.

\item The proposed theory also describes very well the multifractality of the velocity difference, in the sense that, for developed energy spectrum, the scaling exponents of the longitudinal velocity difference, when expressed in terms of the moments order, exhibit the characteric shape observed by the various authors. 
\end{enumerate}

\section{\bf  Acknowledgments}

This work was partially supported by the Italian Ministry for the 
Universities and Scientific and Technological Research (MIUR).

\section{\bf Appendix}

The von K\'arm\'an-Howarth equation gives the evolution in time of the
longitudinal correlation function for isotropic turbulence.
The correlation function of the velocity components is the symmetrical second order tensor
$
\ds R_{i j} ({\bf r}) = \left\langle u_i u_j' \right\rangle 
$, 
where $u_i$ and  $u_j'$ are the velocity components at ${\bf x}$ and 
${\bf x} + {\bf r}$, respectively, being $\bf r$ the separation vector.
The equations for $R_{i j}$ are obtained by the Navier-Stokes equations 
written in the two points ${\bf x}$ and ${\bf x} + {\bf r}$ \cite{Karman38, Batchelor53}.
For isotropic turbulence $R_{i j}$ can be expressed as
\bea
R_{i j} ({\bf r}) = u^2 \left[ (f -g) \frac{r_i r_j}{r^2} + g \delta_{i j}\right] 
\eea
$f$ and $g$ are, respectively, longitudinal and lateral correlation functions, which are
\bea
\ds f(r)= \frac{\left\langle u_r({\bf x}) u_r({\bf x}+{\bf r}) \right\rangle }{u^2}, \
\ds g(r)= \frac{\left\langle u_n({\bf r}) u_n({\bf x}+{\bf r}) \right\rangle }{u^2}
\eea
where $u_r$ and $u_n$ are, respectively, the velocity components parallel and normal to  $\bf r$, whereas $r = \vert {\bf r} \vert$ and 
$u^2$ = $\left\langle u_r^2 \right\rangle$ =$\left\langle u_n^2 \right\rangle$=
$1/3 \left\langle u_i u_i \right\rangle $.
Due to the continuity equation, $f$ and $g$ are linked each other by the relationship
\bea
g = f + \frac{1}{2}  \frac{\partial f}{\partial r} r
\label{g}
\eea

The von K\'arm\'an-Howarth equation reads as follows \cite{Karman38, Batchelor53}
\bea
\ds \frac{\partial u^2 f}{\partial t} = 
\ds  K  +
\ds 2 \nu u^2  \left(  \frac{\partial^2 f} {\partial r^2} +
\ds \frac{4}{r} \frac{\partial f}{\partial r}  \right) 
\label{vk}  
\eea
where $K$ is an even function of $r$, which is defined by the following equation \cite{Karman38, Batchelor53}
\bea
\left(    r \frac{\partial}{\partial r}  + 3  \right) K(r) =
\ds  \frac{\partial }{\partial r_k} 
\ds  \left\langle u_i  u_i' (u_k - u_k')  \right\rangle 
\label{vk1}  
\eea
and which can also be expressed as
\bea
K(r)= u^3 \left(  \frac{\partial}{\partial r}  + \frac{4}{r}  \right) k(r)
\label{kk}
\eea
where $k$ is the longitudinal triple correlation function
\bea
\ds k(r)= \frac{\left\langle u_r^2({\bf x}) u_r({\bf x}+{\bf r}) \right\rangle }{u^3}
\eea

The boundary conditions of Eq. (\ref{vk}) are \cite{Karman38, Batchelor53}
\bea
f(0) = 1, \ \ \lim_{r \rightarrow \infty} f(r) = 0
\label{bc}
\eea
The viscosity is responsible for the decay of the turbulent kinetic energy, the rate of which is obtained putting $r=0$ in the von K\'arm\'an-Howarth equation, i.e.
\bea
\frac{\partial u^2} {\partial t} = 10 \nu u^2 \frac{\partial^2 f}{\partial r^2}(0) 
\label{ke}
\eea
This energy is distributed at different wave-lengths according to the energy spectrum 
$E(\kappa)$ which is calculated as the Fourier Transform of $f u^2$, whereas 
the ''transfer function'' $T(\kappa)$ is the Fourier Transform of $K$ \cite{Batchelor53}, i.e.
\bea
\hspace{-1.0mm}
\left[\begin{array}{c}
\hspace{-1.0mm} \ds E(\kappa) \\\\
\hspace{-1.0mm} \ds T(\kappa)
\end{array}\right]  
\hspace{-1.5mm}= 
\hspace{-1.5mm} \frac{1}{\pi} 
\hspace{-1.0mm} \int_0^{\infty} 
\hspace{-1.5mm}\left[\begin{array}{c}
\hspace{-1.0mm} \ds  u^2 f(r) \\\\
\hspace{-1.0mm} \ds K(r)
\end{array}\right]  \kappa^2 r^2 \hspace{-1.0mm} 
\left( \hspace{-0.5mm}\frac{\sin \kappa r }{\kappa r} - \cos \kappa r \hspace{-0.5mm} \right) d r 
\label{Ek}
\eea
where $\kappa = \vert {\bf \bfkappa} \vert$ and $T(\kappa)$ identically 
satisfies to the integral condition
\bea
\int_0^\infty T(\kappa) d \kappa = 0
\label{tk0}
\eea 
which states that $K$  does not modify the total kinetic energy.
The rate of energy dissipation $\varepsilon$ is calculated for 
isotropic turbulence as follows \cite{Batchelor53}
\bea
\ds \varepsilon = -\frac{3}{2} \frac{\partial u^2} {\partial t}=
  2 \nu \int_0^{\infty} \kappa^2 E(\kappa) d \kappa
\eea
The microscales of Taylor $\lambda_T$, and of Kolmogorov $\ell$, are defined as
\bea
\begin{array}{c@{\hspace{+0.2cm}}l}
\ds \lambda_T^2 = \frac{u^2}{\langle (\partial u_r/\partial r)^2 \rangle} = 
-\frac{1}{\partial^2 f/\partial r^2(0)}, \ 
\ds \ell = \left( \frac{\nu^3} { \varepsilon}\right) ^{1/4}
\end{array}
\eea


\end{document}